\documentclass[fleqn,usenatbib]{mnras}

\usepackage{newtxtext,newtxmath}

\usepackage[T1]{fontenc}

\DeclareRobustCommand{\VAN}[3]{#2}
\let\VANthebibliography\thebibliography
\def\thebibliography{\DeclareRobustCommand{\VAN}[3]{##3}\VANthebibliography}

\usepackage{graphicx}	
\usepackage{amsmath}	

\title[SED modeling of 1ES 1727+502]{Multiwavelength temporal and spectral study of TeV blazar 1ES 1727+502 during 2014 to 2021}

\author[Prince et al.]{
Raj Prince$^{1}$\thanks{E-mail: raj@cft.edu.pl},
Rukaiya Khatoon$^{2,3}$,
Pratik Majumdar$^{4,5}$, 
Bo{\.z}ena Czerny$^{1}$, and
Nayantara Gupta$^{6}$
\\ \\
$^{1}$Center for Theoretical Physics, Polish Academy of Sciences, Al.Lotnikow 32/46, 02-668, Warsaw, Poland\\
$^{2}$Tezpur University, Napaam-784028, Assam, India\\
$^{3}$Inter-University Center for Astronomy and Astrophysics, Post Bag 4, Ganeshkhind, Pune-411007, India\\
$^{4}$Saha Institute of Nuclear Physics, HBNI, Kolkata, West Bengal 700064, India\\
$^{5}$Faculty of Physics and Applied Informatics, Department of Astrophysics, University of Lodz, PL-90236, Poland\\
$^{6}$Raman Research Institute, Sadashivanagar, Bangalore 560080, India
}

\date{Accepted XXX. Received YYY; in original form ZZZ}

\pubyear{2021}

\begin{document}
\label{firstpage}
\pagerange{\pageref{firstpage}--\pageref{lastpage}}
\maketitle

\begin{abstract}
One of the most important questions in blazar physics is the origin of broadband emission and fast-flux variation. In this work, we studied the broadband temporal and spectral properties of a TeV blazar 1ES 1727+502 and explore the one-zone synchrotron-self Compton (SSC) model to fit the broadband spectral energy distribution (SED).
We collected the long-term (2014-2021) multiband data which includes both the low and high flux states of the source.
The entire light curve is divided into three segments of different flux states  and the best-fit parameters obtained by broadband SED modeling corresponding to three flux states were then compared.
The TeV blazar 1ES 1727+502 has been observed to show the brightest flaring episode in X-ray followed by optical-UV and gamma-ray.
The fractional variability estimated during various segments behaves differently in multiple wavebands, suggesting a complex nature of emission in this source. This source has shown a range of variability time from days scale to month scale during this long period of observations between 2014-2021.
A "harder-when-brighter" trend is not prominent in X-ray but seen in optical-UV and an opposite trend is observed in $\gamma$-ray. The complex nature of correlation among various bands is observed. The SED modeling suggests that the one-zone SSC emission model can reproduce the broadband spectrum in the energy range from optical-UV to very high energy $\gamma$-ray.

\end{abstract}




\section{Introduction}
Active galactic nuclei (AGN) have been discovered a century ago and still remain a hot topic of research. The theoretical predictions and the observational results suggest that the AGN have three main components namely, a central super massive black hole (SMBH), an accretion disk around the SMBH, bipolar high relativistic jets of particles. The launching of the jets perpendicular to the accretion disk plane remains a long standing puzzle to the astronomers. 
The GRMHD simulations have made some progress in understanding the launching of the jets and it is suggested to be the interplay between accretion flow, magnetic field lines and the SMBH (\citealt{Monika2013, Monika2014}). Simulations and theory together suggest that the magnetic field plays an important role in collimating the particles and giving them a jet like shape (\citealt{Blandford1979, Blandford1982}, \citealt{Koide1998}, \citealt{Monika2016}). The jets are believed to be highly relativistic and the process of particle acceleration inside the jets are still unclear. The AGNs are classified in various types under the AGN unification scheme developed by \citet{Urry_1995}. According to this scheme, the observer's viewing angle with respect to jet axis divides the sources in various class such as quasars, seyfert galaxies, blazars, etc. In case of blazars, the jet axis is within few degrees to the observer's line of sight. Due to the direct view of the jet and because of the superluminal motion, blazars are among the brightest sources in the Universe. Blazars are classified in two main types namely flat spectrum radio quasars (FSRQ) and the BL Lacertae (BL Lac) objects depending upon the presence or absence of optical emission lines in their spectra (\citealt{Stickel1991}, \citealt{Weymann1991}).
Based on the location of the synchrotron peak, the BL Lac objects are further divided into three parts known as Low BL Lac (LBL), intermediate BL Lac (IBL), and high BL Lac (HBL) by \citet{Padovani1995}. 

The main characteristics of blazars are the following, they show stochastic flux variability and strong flaring activity when they are very bright and subsequently come down to the normal flux states. The period of flaring activity can be as low as one day and it can stay in a high state for a long period of time, even up to years.
Blazars have been observed across all the wavebands by various space-based and ground-based telescopes. 
 The total spectral energy distribution (SED) of a blazar covers a wide range in frequency starting from radio to very high energy (VHE) $\gamma$-ray. The shape of the SED has two broad hump structure one peaking in low energy band ranging from radio to X-ray and other hump peaks in high energy regime covering the energy range from X-ray to VHE gamma-ray. 
The emission from radio to soft X-ray has been explained by the synchrotron emission produced by the motion of relativistic electrons in jet magnetic field (\citealt{Abdo_2010}). The observations in hard X-rays and in gamma-rays are currently expected to be produced by the inverse-Compton scattering of internal (synchrotron-self Compton or SSC; \citealt{Konigl1981}, \citealt{Marsche1985}, \citealt{Ghisellini1989}) and external low energy photons by the highly relativistic electrons (external Compton or EC; \citealt{Begelman1987}) generally termed as leptonic scenario. The source of external seed photons could be the accretion disk (\citealt{Dermer1993}, \citealt{Boettcher1997}), the broad-line region or BLR (\citealt{Sikora1994}, \citealt{Ghisellini1996}), and the dusty torus or DT (\citealt{Baejowski2000}, \citealt{Ghisellini2008}).
Apart from the leptonic model, the high energy peak has been explained by hadronic (\citealt{MUCKE2003}, \citealt{Bottcher2013}) and lepto-hadronic (\citealt{Diltz_2016}, \citealt{Paliya_2016}) model as well in many  blazars. In hadronic model, the high energy peak of spectral energy distribution (SED) is produced by the proton-photon or proton-proton interactions inside the jets. In the past studies, the leptonic/hadronic models have been extensively used to explain the broadband SEDs of blazars. 

The flaring period can vary depending upon the cause of the flare and in leptonic scenario it is expected to be flaring in all wavebands simultaneously. But, we also sometimes observe a time delay between emissions in different wavebands, which challenges the standard one-zone leptonic emission model (\citealt{2015MNRAS.448.3121H}, \citealt{Paliya_2016},\citealt{Cerruti2017},\citealt{Xue2019},\citealt{Prince2019b}, \citealt{Avik2021}, \citealt{2022A&A...657A..20B}). The fast flux variability is another important property of these sources. 
Many blazars have shown a range of variability timescale extending from minutes to several days
(\citealt{Heidt1996}, \citealt{Ulrich1997}, \citealt{Ackermann2016},\citealt{Prince2019b},\citealt{Chatterjee2021}). It has been suggested that the observed variability time physically indicates the size of the region where the emission is produced and in minute scale variability it is always challenging to explain the broadband emission produced close to the SMBH in a relatively small region. Apart from that the underlying physical mechanism of accelerating charged particles up to relativistic speed is hard to explain, currently the shock acceleration and the magnetic re-connection (for a review see \citealt{Blandford2019}) are the best models to explain acceleration of charged particles. 
The fractional variability is used to derive the degree of variability in blazar light curves and recently, a tool has been developed by \citet{galaxies7020062} taking full consideration of systematic effects and the light curve sampling rate across the wavebands. They found that the variability in TeV range for blazars Mrk421 and Mrk 501 are higher than the previously reported values. Recently, more detailed study of long-term variability is done by \citet{Polkas2021} and \citet{Thiersen_2022} where they try to explain it by the numerical simulations.

The blazar, 1ES 1727+502 (z = 0.055, \citealt{Vaucouleurs1991}) is identified as a BL Lac type source and later it was also observed in very high energy gamma-ray by MAGIC telescope (\citealt{Aleksi_2011, Aleksic2014}) and hence known as TeV blazar. Again in 2013, very high energy photons above 250 GeV have been observed with VERITAS (\citealt{Archambault2015}) confirming the TeV nature of the source.
Our study focuses on its recent X-ray flare, followed by in optical-UV and eventually in gamma-ray. Very high energy photons above 300 GeV is also detected by MAGIC (\citealt{MAGIC2020}) during the same period.
Though the source is not very bright in Fermi-LAT energy band, simultaneous flaring was observed in X-ray and gamma-ray band. Two-zone model has been used previously to fit the broadband SED of 1ES 1727+502 (from radio to very high energy gamma-ray) along with a few others BL Lac objects in \citet{MAGIC2020}. In this work, we do not have radio data and have shown that one-zone SSC model is enough to explain the broadband SED of 1ES 1727+502 during various flux states.
We have collected broadband data of this source and a detailed temporal and spectral study is done to answer a few of the open questions on blazars.

We arrange our study in following way: Data analysis is presented in section 2, section 3 discusses the temporal behavior, broadband correlation is presented in section 4, followed by color-magnitude variation in section 5 and SED modeling in section 6, summary and discussions is presented in section 7, and eventually conclusions in section 8.

\section{Data reduction and Analysis}
\subsection{Fermi-LAT}
Fermi satellite was launched by NASA in July 2008 with two gamma-ray detectors. The first detector is large area telescope (LAT) which is mainly designed to scan the entire sky in gamma-ray energy band 20 MeV to 300 GeV and search for the transient gamma-ray sources, while the second one is called gamma-ray burst monitor (GBM) mainly dedicated to catch the gamma-ray bursts. The LAT has observed more than five thousands extra-galactic 
gamma-ray sources (4FGL; \citealt{Abdollahi_2020}) from the continuous monitoring between 2008-2018, among them more than three thousands objects are blazar, which suggests that the gamma-ray sky is highly populated by relativistic jets. LAT has a FoV of 2.4 Sr (\citealt{Atwood2009}), which covers 20$\%$ of the sky at any time. The LAT orbit is such that it can scan the entire sky in every three hours. We have analysed the LAT data from December 2014 (MJD 57000) to March 2021 (MJD 59300) following the standard Fermi tutorials\footnote{\href{https://fermi.gsfc.nasa.gov/ssc/data/analysis/scitools/}{Fermi tools}}.
Standard selection criteria for the tools \texttt{gtselect}, \texttt{gtmktime}, \texttt{gtltcube}, and \texttt{gtexposuremap} are imposed based on the suggestion of Fermi science support cell. As suggested by science team, we have used evclass=128 for the photon like events and evtype=3 for all types of events (front+back). Further the analysis is performed with the pass 8 data.
The secondary gamma-ray photons from the earth limb are suppressed by choosing the maximum zenith angle 90$^\circ$. The latest galactic background (gll\_iem\_v07) and the isotropic galactic emission (iso\_P8R3\_SOURCE\_V2\_v1) models have been used to properly model the galactic as well as extra-galactic gamma-ray background. The model and the parameters of the sources within 10$^\circ$ of region of interest (ROI) are used from the recently published 4FGL-DR2 (\citealt{Abdollahi_2020}) catalog. Initially, all the sources within 10$^\circ$ are kept free and their spectral parameters are optimized with the maximum Likelihood procedure. 


\subsection{Neil Gehrels Swift Observatory}
Neil Gehrels Swift Observatory is a space-based observatory, which carries three main instruments namely burst alert telescope (BAT; 3.0-150.0 keV), X-ray focusing telescope (XRT; 0.3-10.0 keV), and UV optical telescope (UVOT; 170 - 650 nm) with the main objective to solve the origin of gamma-ray bursts. Soon after its launch, the instruments started looking for the transient activity in X-rays, ultraviolet (UV) and optical bands. 
The main advantage of having the Swift telescope is, it provides true simultaneous data in all wavebands ranging from X-ray to optical,
which is crucial to understand the physical processes involved in the emission. Swift has observed many Fermi blazars simultaneously in all wavebands. We collected all the XRT and UVOT data available for the blazar 1ES 1727+502. 

{\bf XRT:}
The standard data reduction procedure\footnote{https://www.swift.ac.uk/analysis/xrt/} has been followed to extract the source and background counts from the circular region of 10$''$ and 30$''$ radius around the source and away from the source respectively. Source and background spectrum were created with the help of \texttt{XSELECT}, which were
used in \texttt{XSPEC} for the modeling. The generated spectrum was grouped with the ancillary response file (ARF) and the redistribution matrix files, using the tool \texttt{grppha}. The spectra were grouped with 20 counts per bin to have sufficient counts in each bin.
The resultant grouped spectrum is then loaded in \texttt{XSPEC} and fitted with  a simple unabsorbed powerlaw spectral model .
We fitted each observation during December 2014 to March 2021 with the above model and estimated the flux and the corresponding power-law index. The interstellar absorption has been corrected with the HI column density, n$_{\rm H}$ = 0.02$\times$10$^{22}$ cm$^{-2}$.

{\bf UVOT:} 
The observations performed in three optical filters ( e.g., U, B, and V) and three UV filters (e.g., W1, M2, \& W2) of UVOT are used in our study. 
In this case, the source and background regions were selected with equal radius 5$''$ around the source and way from the source respectively.
The task \texttt{UVOTSOURCE} is used to extract the source and background magnitude which are further corrected for the galactic reddening (E(B-V)=0.079; \citealt{Schlafly_2011}) and the atmospheric extinction (A$_{\lambda}$). Further, the magnitudes were converted to fluxes following the procedure by \citet{Breeveld_2011} and \citet{Giommi2006}.

\section{Broadband light curves}
\begin{figure*}
\centering
\includegraphics[scale=0.4]{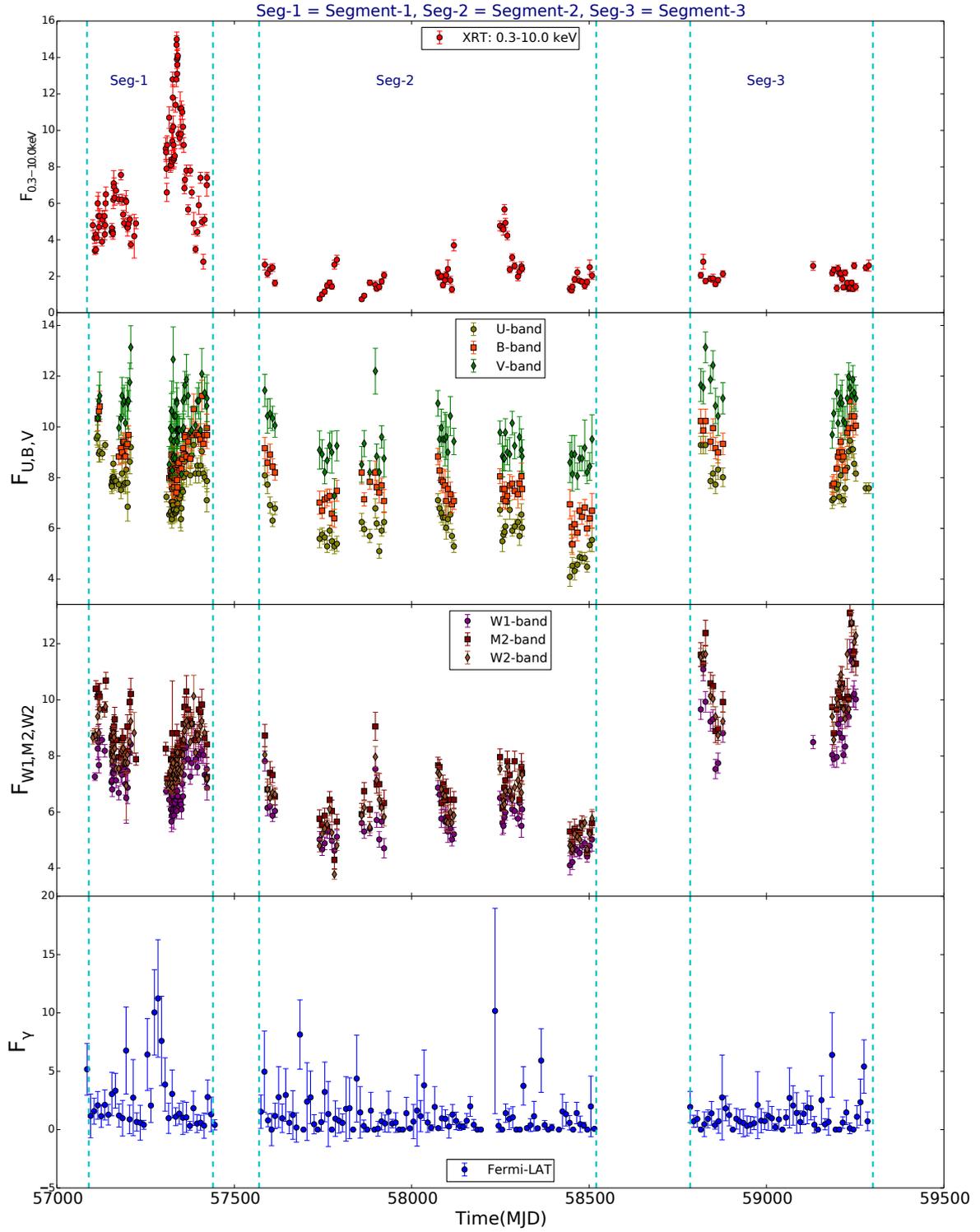}
\caption{The broadband light curves of BL Lac 1ES 1727+502. X-rays are in unit of 10$^{-11}$ erg cm$^{-2}$ s$^{-1}$, optical-UV are in units of 10$^{-12}$ erg cm$^{-2}$ s$^{-1}$, and gamma-rays are in unit of 10$^{-8}$ ph cm$^{-2}$ s$^{-1}$. }
\label{fig1}
\end{figure*}

 The TeV blazar 1ES 1727+502 flared in X-ray band during the year 2015 observed by Swift-XRT and simultaneously a very high energy gamma-ray flare with energy above 300 GeV was detected by MAGIC (\citealt{MAGIC2020}).
Simultaneous data collection in the optical-UV wavebands with Swift-UVOT also recorded a bright state. Continuous monitoring of gamma-ray sky with Fermi-LAT also detected this source with mild flare as expected for TeV blazar. We collected the broadband data from December 2014 to June 2021 to study the long-term behavior of the source. Total light curves are shown in Figure \ref{fig1} and divided into three different segments namely, Segment-1 (MJD 57090--57440), Segment-2 (MJD 57570--58520), and Segment-3 (MJD 58785--59300). 

\begin{figure*}
    \centering
    \includegraphics[scale=0.4]{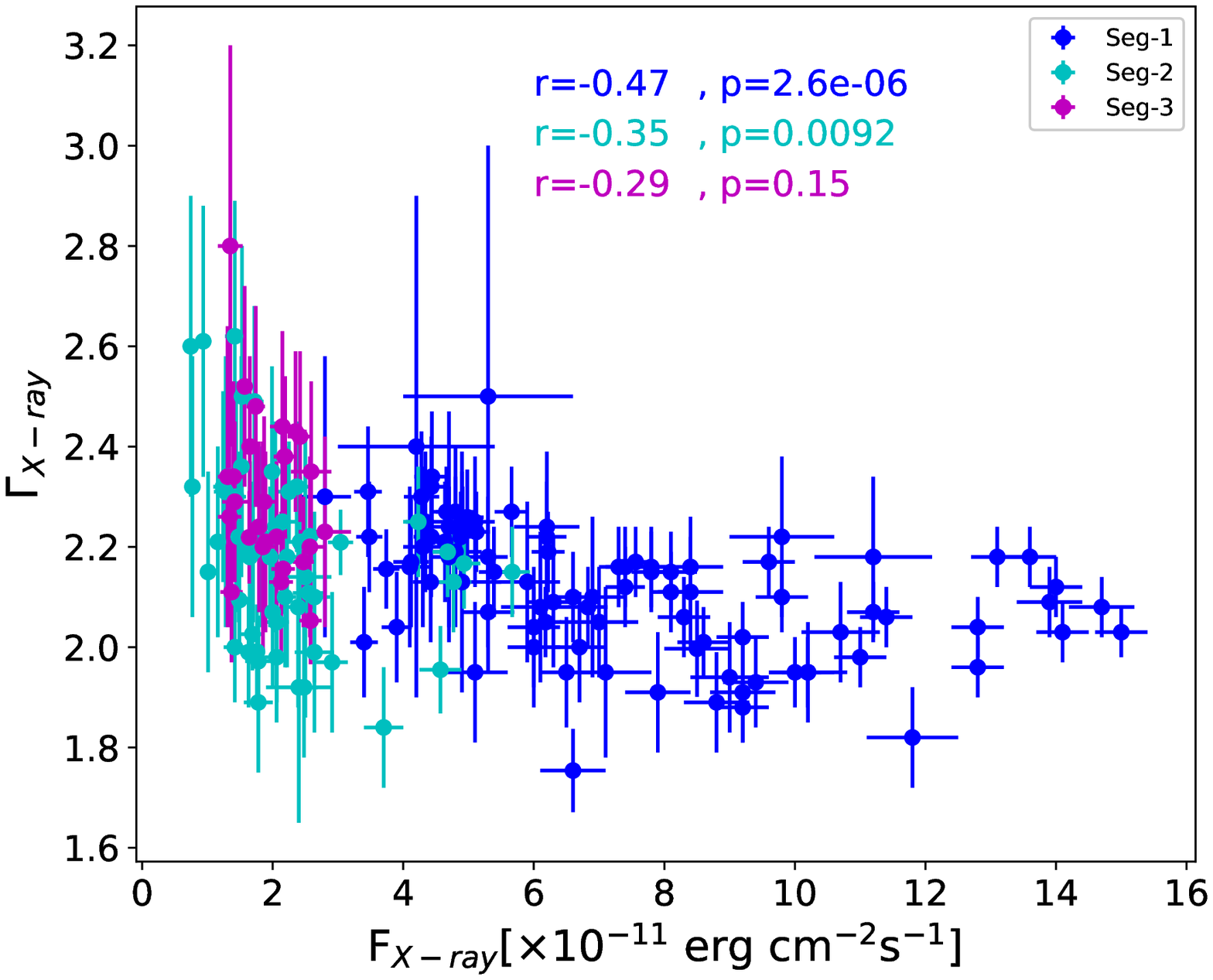}
    \includegraphics[scale=0.54]{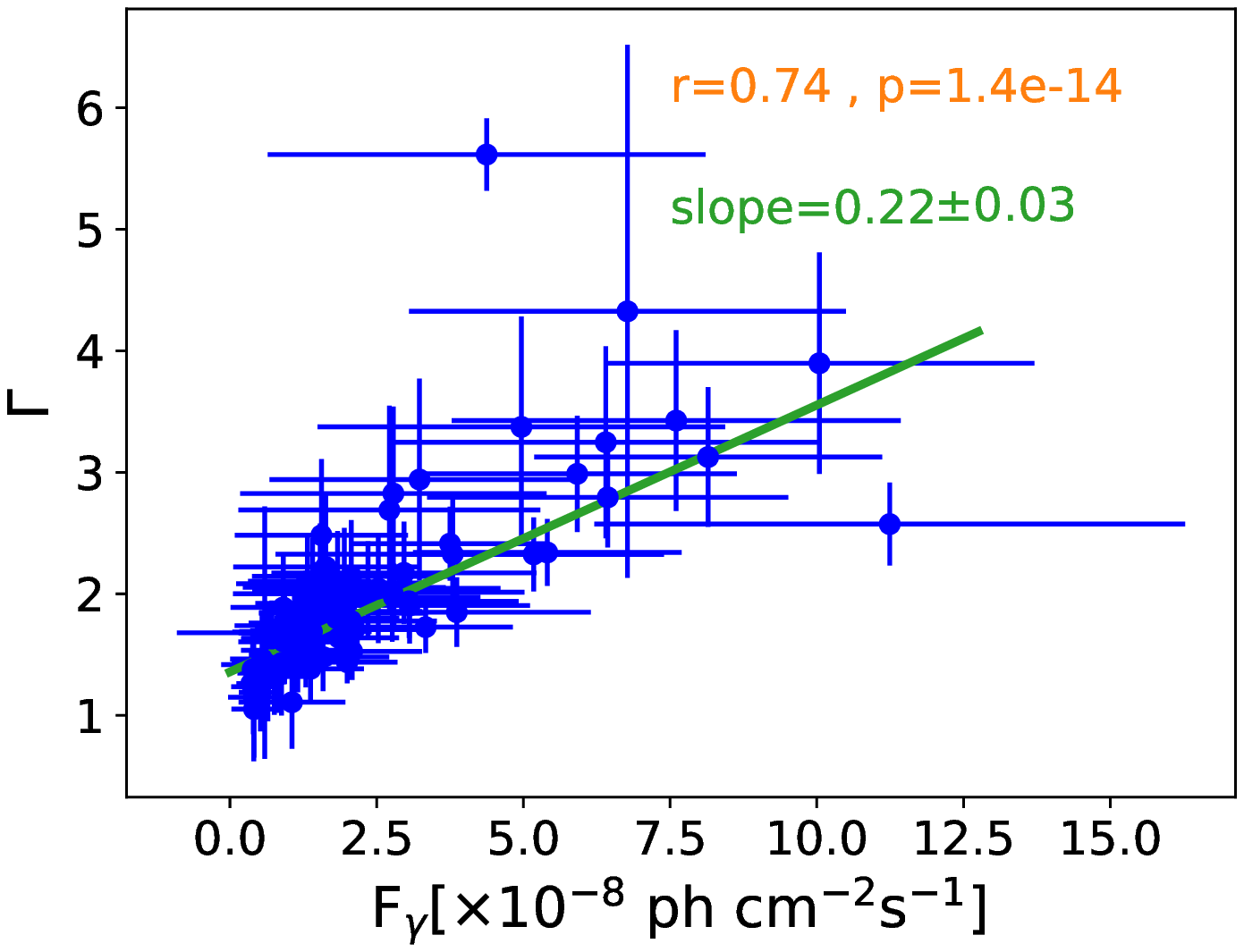}
    \caption{Left panel: X-ray photon spectral index with corresponding fluxes does not show a clear "softer-when-brighter" trend. Right panel: The gamma-ray spectral index plotted against the corresponding fluxes suggests a "softer-when-brighter" behavior.  }
    \label{fig2}
\end{figure*}

In Figure \ref{fig2}, we have shown the spectral variation in X-ray and gamma-ray energy bands with respect to fluxes in respective bands. In X-ray the various segments were separated by a few months and hence we show their spectral behavior separately in different color.
Pearson's correlation coefficient and corresponding null hypothesis are estimated to quantify the level of correlation. 
The correlation results suggest a hint of anti-correlation in flux and index in the X-ray band for all the segments. However, 
in the gamma-ray band they show positive correlation.
 
To visualize the trend in, gamma-ray, the flux vs. index plots, we fitted the data points with a straight line (green bold line in Figure \ref{fig2}).
A "brighter-when-harder" trend is usually seen in X-ray for TeV blazar. However, we do not observe any clear trend in X-ray but rather a hint of "brighter-when-harder" trend. On the other hand, a "softer-when-brighter" trend is seen in gamma-ray. The absence of any particular trend or different trends seen in X-ray and $\gamma$-ray imply that these two emissions are produced by the higher and lower end of the electron population respectively, as also discussed in the last paragraph of Section 3.4.
Several studies suggest that the X-rays are produced via synchrotron process and the gamma-rays are produced by the inverse-Compton scattering of synchrotron photons by the relativistic leptons. 

\subsection{Segment-1}
\begin{figure*}
\centering
\includegraphics[scale=0.34]{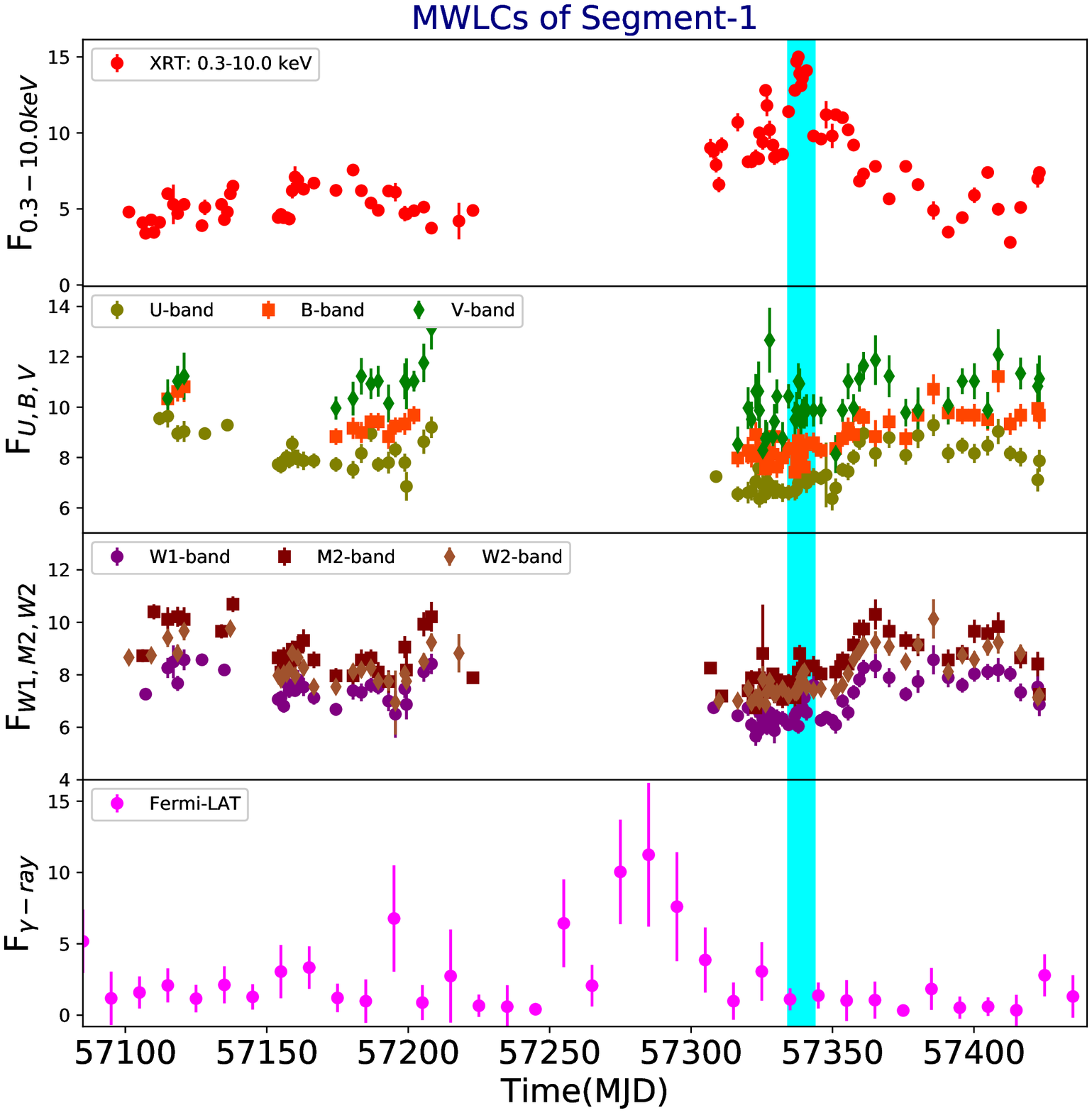}
\includegraphics[scale=0.34]{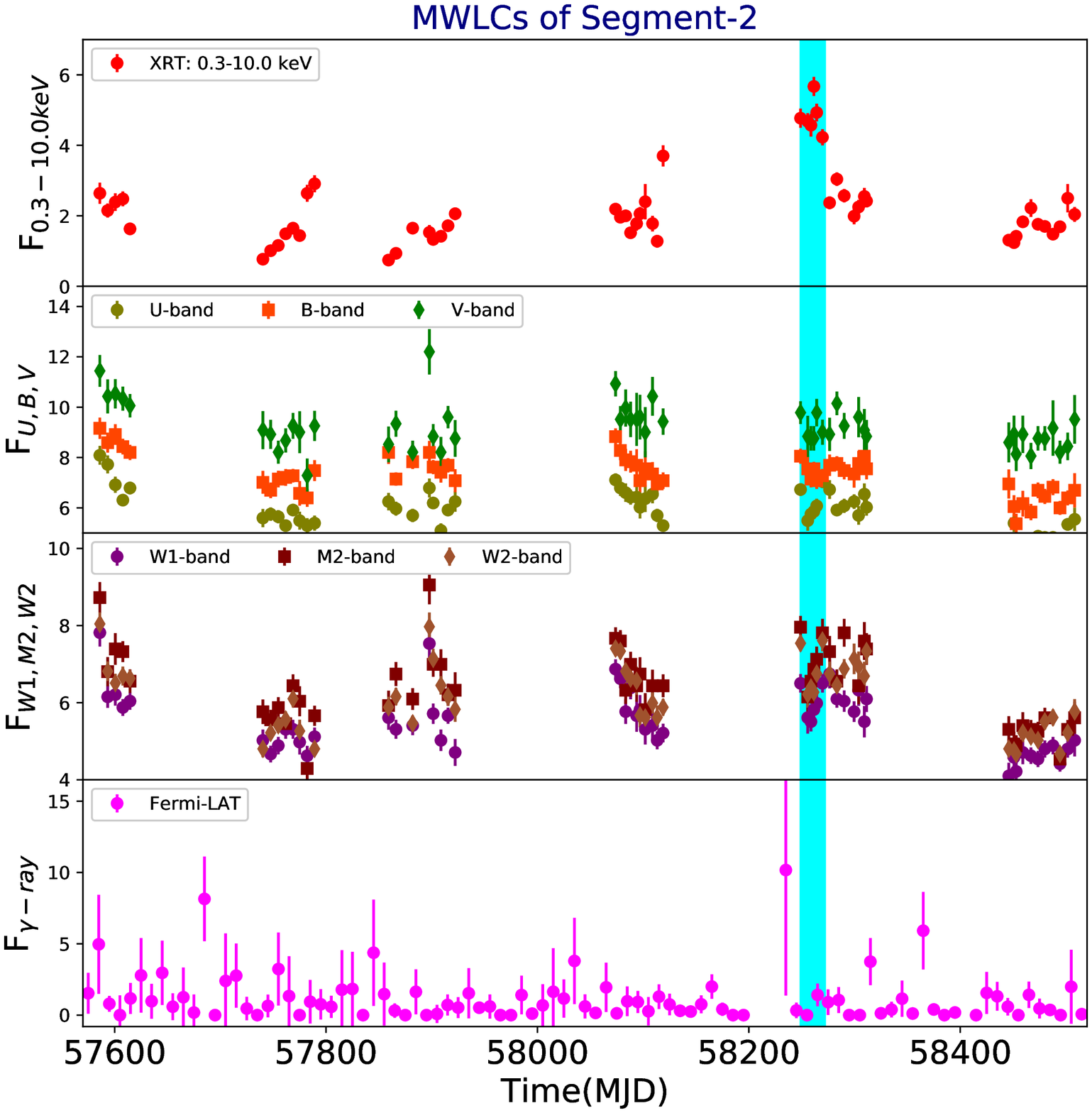}
\includegraphics[scale=0.34]{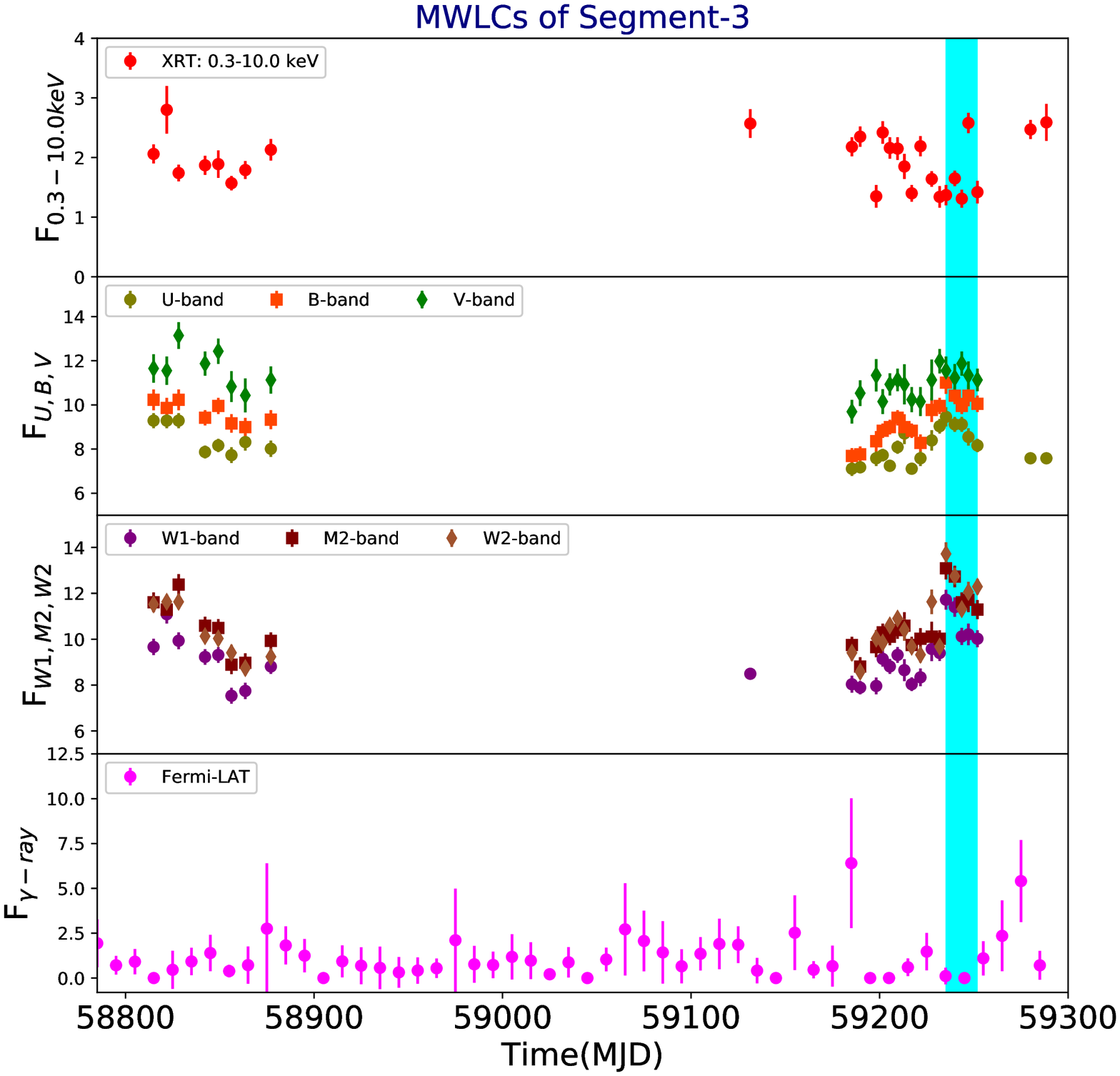}
\caption{Broadband light curves of each segments. The color patch marked the period used for SED modeling. The marked period in Segment-1 is between MJD 57336-57346, in Segment-2 it is between MJD 58248-58273, and in Segment-3 it is between 59235-59252.}
\label{fig3}

\end{figure*}

The good coverage of broadband data in X-ray, optical-UV, and gamma-ray is found to be useful during Segment-1. It is identified as the brightest X-ray flaring period. The source is also quite bright in optical-UV and gamma-ray. The broadband light curves are shown in Figure \ref{fig3}. The time of exact gamma-ray flaring is missed in X-ray and optical-UV, and after almost two months a bright X-ray flare is observed during the low gamma-ray flux states. In the first part of the light curve (MJD 57090--57220), the X-ray flux is quite high above $\sim$5$\times$10$^{-11}$erg cm$^{-2}$s$^{-1}$ and almost constant during this period. However, the fluxes in optical U and V bands are varying with time whereas in B band the flux is almost constant. 
In UV bands the flux appears to be constant for this period. However, under the leptonic scenario all optical-UV should follow the same variability pattern produced by the synchrotron emission. Here the discrepancy suggests a more complex nature of the flaring event. 
On the other hand, the other part of the light curve (MJD 57300--57440) shows a bright peak in X-ray followed by a decay to a lower flux state. A rising trend is seen in optical and UV light curves during this period which is opposite to the X-ray emission. Further, we perform the correlation study between X-ray and optical-UV, in later section, to understand the behavior of the light curves. In both the parts the gamma-ray emission is in low flux state showing a constant behavior. During the Segment-1 between MJD 57307-57327 a photon of energy more than 300 GeV was observed as reported in \citet{MAGIC2020}.

\subsection{Segment-2}
The Segment-2 (MJD 57570--58520) consists of chunks of observations in X-ray, optical and UV bands, while observations are continuous in gamma-ray band, as shown in Figure \ref{fig2}.
The continuous gamma-ray data shows a constant behavior with low flux state below $\sim$2.0$\times$10$^{-7}$ ph cm$^{-2}$s$^{-1}$. The X-ray light curve shows a long-term variability where flux rises above $\sim$1$\times$10$^{-11}$erg cm$^{-2}$s$^{-1}$ and reaches upto $\sim$6$\times$10$^{-11}$erg cm$^{-2}$s$^{-1}$ with each chunk showing a small scale variability within a few days. A small scale variability (on shorter scale) is also detected in optical and UV bands. We note that the highest fluxes observed in X-ray, optical and UV bands during Segment-2 are below the fluxes observed for Segment-1. The variability study is discussed in detail in section 3.4. 

\subsection{Segment-3}
The Segment-3 (MJD 58785--59300) consists of two chunks of observations separated by a few months.
The X-ray flux in this segment is lower compared to the fluxes in Segment-1 and 2, and almost constant over a longer period of time. A short time scale variability is present and examined in detail in later section. The source is also observed to be in low flux state in gamma-rays during this period with an almost constant flux. However, the source is brighter in optical-UV suggesting an orphan flare.
Orphan flares are rarely seen but they have been observed in a few sources (\citealt{Rajput2019,Rajput2020}) and mostly seen in optical-UV and IR bands (\citealt{Chatterjee2013}). Detailed temporal and spectral study of Segment-3 and comparison with Segment-1 and Segment-2 can answer the reason behind the orphan flare. A significant flux variability is observed in optical-UV bands and the flare is as strong as seen in Segment-1 with maximum V and W2 band fluxes close to $\sim$1.4$\times$10$^{-11}$ erg cm$^{-2}$ s$^{-1}$. Flare in UV is even brighter than the earlier flares seen during Segment-1.

\subsection{Characteristic Variability}
\begin{figure*}
    \centering
    \includegraphics[scale=0.46]{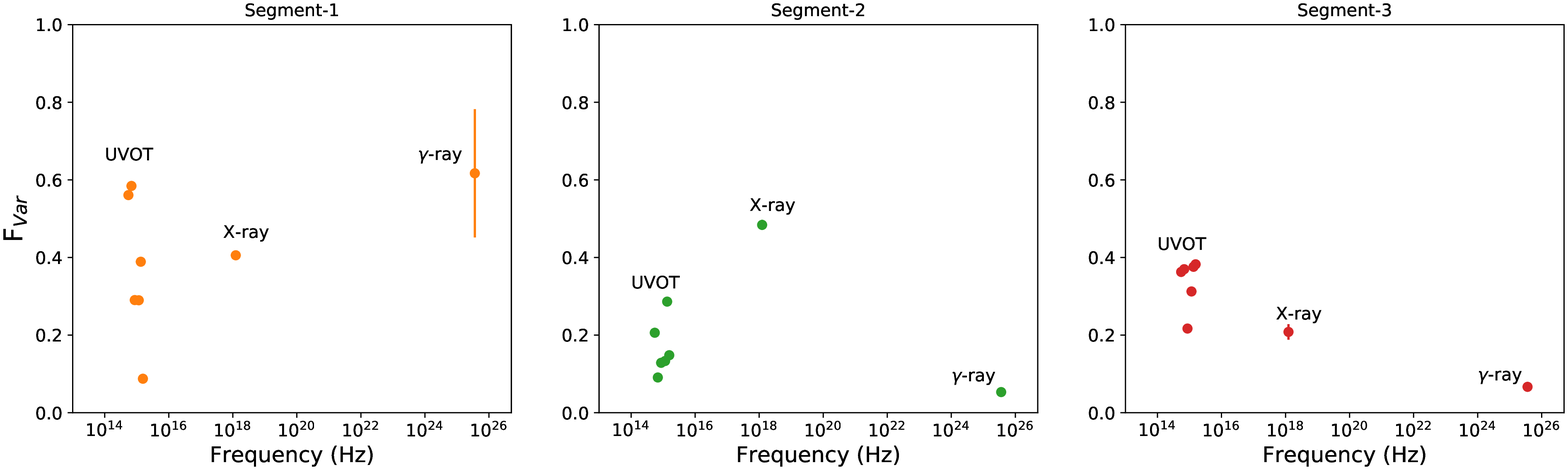}
    \caption{Fractional variability estimated during various segments for different wavebands, suggesting variability is more in segment-1 compared to segment-2 and 3.}
    \label{fig:fvar}
\end{figure*}

To characterize the variability of the source during the various segments across the wavebands, we estimated the fractional variability amplitude and the fastest variability time. 
The fractional variability amplitude tells about the long-term variability present in the source, formulated by \citet{Vaughan_2003} as:

\begin{equation}
 F_{\rm var} = \sqrt{\frac{S^2 - err^2}{F^2}},
\end{equation}
where, F is the mean flux, S$^2$ and err$^2$ are the variance and the mean square error in the flux, respectively. The formula to estimate the error in fractional variability amplitude is provided in \citet{Prince2019a}. The F$_{\rm var}$ estimated for all the segments at various wavebands are depicted in Table \ref{tab:Fvar}. The plot of F$_{\rm var}$ vs. frequency is shown in Figure \ref{fig:fvar} to visualize the shape of it.
Surprisingly, all the segments behave differently suggesting a complex nature of the source. During Segment-1, $\gamma$-ray shows highest variability followed by optical B \& V bands and then X-ray and finally to UV bands. Similar variability pattern is also seen in many other blazars (\citealt{Rani2017, Rani2018})
Segment-2 recorded the highest variability in X-ray band followed by optical-UV and then $\gamma$-ray. A similar result is also obtained for the other blazar by \citet{2011ApJ...736..131A, 2016ApJ...819..156B}.
Segment-3, in which optical-UV orphan flare is identified, shows the maximum variability in optical-UV followed by X-ray and then $\gamma$-ray . This behavior is unique and has never been seen in any blazar before.
In general, most of the blazar the fractional variability resembles the broadband SED (\citealt{2015A&A...576A.126A,2015A&A...578A..22A}) which is also seen for segment-1 and segment-2.

The flux variability time is also estimated for all the segments in various wavebands. These values are further used to characterize the variability in the blazar 1ES 1727+502. 
The flux variability time, $t_{var}$, can be defined as (\citealt{Brown2013}),
\begin{equation}
 F_2 = F_1 ~2^{(t_2-t_1)/t_{var}}
\end{equation}
where, F$_1$, and F$_2$ are fluxes measured at time t$_1$ and t$_2$. The variability time or the fastest (or shortest) variability time (t$_{\rm var}$) considered as the smallest value among the available pairs in the entire light curve. During the Segment-1, we found t$_{\rm var}$ of 1-day across the entire wavebands.
However, during Segment-2 \& 3 it varies from days to month across the wavebands, the estimated values are shown in Table \ref{tab:Fvar}. It suggests that the source has both kinds of variability from shorter time scale (day) to longer time scale (months), which is mostly seen in blazars. The variability amplitude and the variability time together can characterize the variability of the source.

The variability time is shorter in X-ray waveband and longer at optical-UV and $\gamma$-ray, suggesting that the X-rays are produced by the highest energy tail of the electron population via synchrotron mechanism. This is the assumption generally made in the context of synchrotron emission, which also implies that the variability at higher energy (gamma-ray) is produced by the lower energy part of the electron population. Similar behavior in variability timescale is also seen for Mrk 421 (\citealt{Ritaban2021}), a TeV blazar.   

Considering the observed variability time in X-ray is caused by the synchrotron cooling, we can estimate the other jet parameters using the expression derived in \citet{Rybicki1979},
\begin{equation}
    t_{cool} \simeq 7.74\times10^8  \frac{(1+z)}{\delta}B^{-2}\gamma^{-1} ~{\rm sec}.
\end{equation}
Where, B is the strength of magnetic field in Gauss and t$_{cool}$ is the synchrotron cooling timescale in seconds. For a Doppler factor, $\delta$ = 29.0, and $\gamma$ = 1.5$\times$10$^{5}$, we estimate the value of the magnetic field to be between 0.046 Gauss (for X-ray variability time 1 day) and 0.025 Gauss (for variability time 3.63 days). The value of the magnetic field optimized in the broadband SED modelling (Table \ref{tab:sed_param} ) is consistent with these derived values.

\begin{table}
\centering
\caption{The broadband fractional variability and the variability time during various segments of period.}
\begin{tabular}{cccc}
\hline \hline
Segments&Wavebands &F$_{\rm Var}$($\%$)  & t$_{\rm var}$ (days)\\
  \hline \hline
Seg-1  & X-ray   & 40.57$\pm$0.63 & 1.00\\
  & U   & 29.03$\pm$0.59 & 1.07\\
  & B    & 58.44$\pm$0.69 & 4.04\\
  & V   &  56.08$\pm$0.95  & 1.77 \\
  & W1  &  28.98$\pm$0.60  & 1.69\\
  & M2  &  38.91$\pm$0.66  & 1.85\\
  & W2  &  8.76$\pm$0.52  & 5.16 \\
  & $\gamma$-ray & 61.70$\pm$16.52 & 2.52\\
  \hline
 Seg-2  & X-ray   & 48.40$\pm$1.29 & 3.63\\
  & U   & 12.86$\pm$0.75 & 8.16\\
  & B    & 9.08$\pm$0.82 & 26.66\\
  & V   &  20.61$\pm$1.01  & 7.83 \\
  & W1  &  13.33$\pm$0.74  & 9.13\\
  & M2  &  28.63$\pm$0.75  & 18.22\\
  & W2  &  14.82$\pm$0.59  &  14.99\\
  & $\gamma$-ray & 5.32$\pm$U.L. & 2.56\\ 
 \hline 
   Seg-3  & X-ray   & 20.82$\pm$2.00 &3.61 \\
  & U   & 21.69$\pm$0.81 & 13.34\\
  & B    & 36.99$\pm$0.90 & 25.00\\
  & V   &  36.28$\pm$1.17  & 21.51 \\
  & W1  &  31.25$\pm$0.81  & 10.61\\
  & M2  &  37.60$\pm$0.85  & 8.77 \\
  & W2  &  38.25$\pm$0.69  &  6.70\\
  & $\gamma$-ray & 6.67$\pm$U.L. &5.42 \\ 
  \hline
\end{tabular}
\label{tab:Fvar}
\end{table}

\section{Correlation study}
The correlation studies among various waveband emissions are important to understand the exact location of their production sites. Under the leptonic scenario it is strongly believed that the broadband emissions (radio to gamma-rays) are produced by a single emitting zone. However, in few cases two-zone emission model is recommended based on the time delay or the different variability observed in various wave bands ( \citealt{Patel2018}, \citealt{Prince2019b}, \citealt{Avik2021}). Along with the location sites it can also help to understand the various physical processes occurring at various locations. The only requirement is to have a very good coverage of data across the wavebands. In the current study, Segment-1 has a dense pointings of Swift to monitor the source. However, the data samples are concentrated on two parts (part-1 \& part-2) separated by more than two months. We cross-correlate the X-ray flux with the optical-UV separately for both the parts. We have used the discrete cross-correlation (DCF) method formulated by \citet{Edelson1988}, the expression can be found in \citet{Polkas2021}.

\begin{figure*}
    \centering
    \includegraphics[scale=0.38]{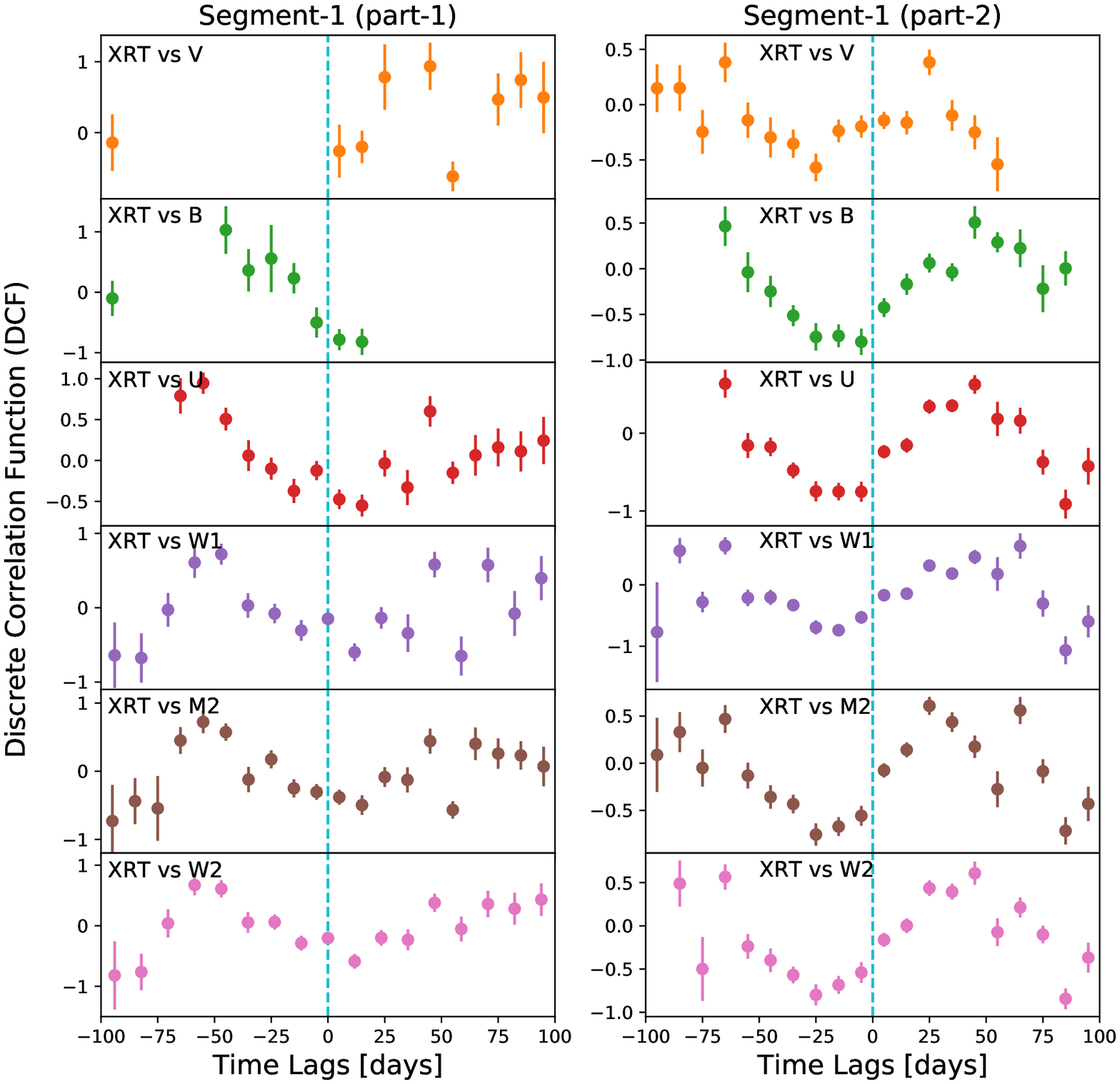}
    \includegraphics[scale=0.38]{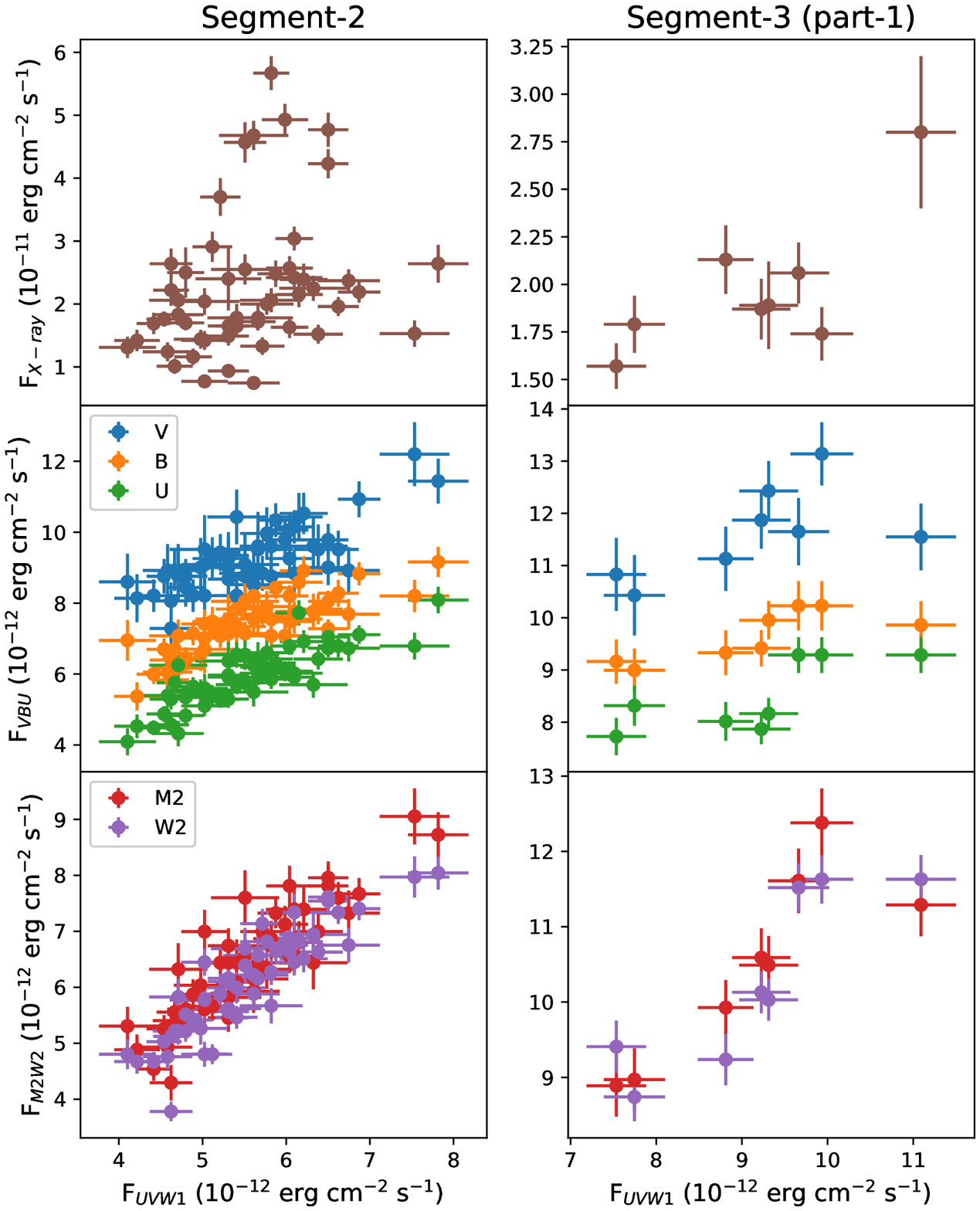}
    \caption{Left panel: DCF estimated for the Segment-1, which is divided into two parts since they are separated by more than two months. Right panel: Linear correlation between the Swift-XRT observations versus all the UVOT bands.}
    \label{fig4}
\end{figure*}

In the part-1, for the B and V bands the DCF is not properly constructed due to the less number of observations. However, other bands (U, W1, M2, and W2) show a strong cross-correlation with X-ray emission at around -50 days, suggesting optical U and ultra-violet (UV) band emissions are leading the X-ray emission. An anti-correlation with $\sim$100 days is also noted but it appeared at the end of the DCF. Hence, it can not be considered as a genuine time lag. This trend can also be confirmed from the multi-wavelength light curves of Segment-1, where optical and UV are in the high flux state in the beginning and X-ray is in low state. As time passes X-ray flux goes up and at the end goes down and opposite trend is seen in optical-UV with going down at the beginning and up at the end. The peak of X-ray ($\sim$ MJD 57160) and the peak of optical-UV (MJD 57210) are approximately separated by 50 days. The DCF plots are shown in Figure \ref{fig4}. Similar result was also reported by \citet{Urry1997}, where they studied the time lag between X-ray and UV for BL Lac PKS 2155-304, and found that X-ray flare comes after the UV flare. Assuming X-ray and optical/UV photons are produced by the synchrotron process, they suggested that  stochastic acceleration process, like diffusive acceleration at shock front, could be responsible for this feature in flare data.

Part-2 has a quite dense sampled data in all the wavebands, and hence a well sampled DCF is produced (Figure \ref{fig4}). The cross-correlation between X-ray and optical-UV emissions shows a significant positive correlation (with correlation coefficient above 50\%) with time lag of 25 days in V and M2 bands with respect to X-ray emission. The time lag of $\sim$50 days is detected between the X-ray emission and the emissions in other bands (B, U, W1, \& W2).
The correlation coefficient at these two time lags are above 50$\%$, suggesting a strong positive correlation.  

Segment-2 covers a quite long-period of interval from MJD 57570 to 58520 ($\sim$2.6 years) and have X-ray and optical-UV observations at occasional periods. In this segment, there are two chunks of observations separated by a few months and hence not sufficient for the correlation study (Figure \ref{fig3}).
Though the gamma-ray observations are continuous but no significant activity is observed. Similarly, during Segment-3 there are two chunks of observations in X-ray, optical and UV bands but separated by almost a year.
We divide the Segment-3 into two parts namely, part-1 and part-2 (see Figure \ref{fig3}).
Due to the less number of observations in part-1, it is not advisable to do the cross-correlation study. As suggested by \citealt{2013arXiv1302.1508A} minimum 11 data points are required to measure the correlation between two bands. To look for linear correlation between optical-UV and X-ray emissions, we plotted their fluxes in the 2-D plot in Figure \ref{fig4}. Optical (U, B, and V) and UV (M2 and W2) show a strong correlation with UVW1 band in both the Segment-2 and part-1 of Segment-3 with zero time lag. However, the correlation between the X-ray emission and the UVW1 band emission shows a non-linear trend, suggesting a complex correlation between them. The flux-flux relation is very common in AGN where the emissions at various wavelength are correlated and show a linear trend (\citealt{2020ApJ...896....1C}), and any dispersion from the linearity leads to a complex nature of the source emission.

\begin{figure}
    \centering
\includegraphics[scale=0.45]{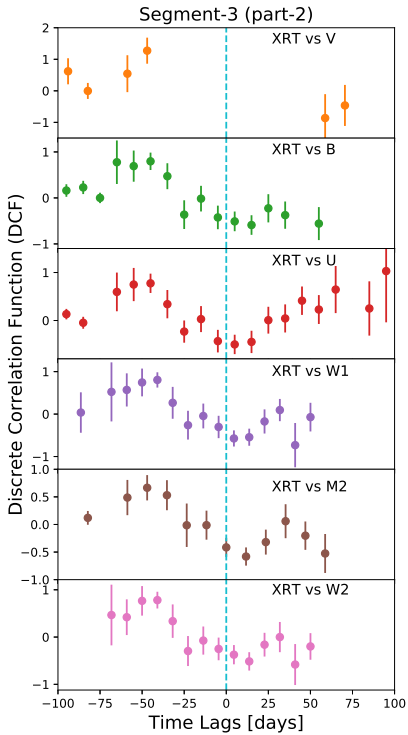}
    \caption{Cross-correlations amongst various wavebands during part-2 of Segment-3. }
    \label{fig5}
\end{figure}

Part-2 light curve of Segment-3 has sufficient number of observations and the DCF analysis is performed to estimate the time lags for the various combinations between X-ray and optical-UV emissions. The DCF results are shown in Figure \ref{fig5}, and a negative time lag of $\sim$50 days is observed between X-ray and optical-UV except with the V band, suggesting X-ray emission lags behind the optical-UV emissions. The correlation coefficients are quite significant with DCF between 60--90$\%$. The time lag of $\sim$50 days between X-ray and optical-UV are also visible in light curves (see Figure \ref{fig3}), X-ray peaks when flux is low in optical-UV and vice versa.

Part-2 of Segment-1 shows a positive time lag of $\sim$50 days between X-ray and optical-UV emissions. As suggested by the \citet{Urry1997}, the time lag at lower frequency with its corresponding synchrotron emission at higher frequency roughly would be the radiative cooling time of relativistic electrons at lower frequency, and the main cooling process is the synchrotron cooling. Then the time lags between the optical/UV and X-ray emission in the observer frame can be defined as (\citealt{Bai2003});
\begin{equation}
    t_{lag} \approx 1.477 \times 10^{4} \frac{(1+z)}{\delta}^{-1/2} B^{-3/2} seconds, 
\end{equation}
Using the magnetic field, B=0.02 (from the SED modeling result, see Table \ref{tab:sed_param}), and $\delta$ =29, we estimated the time lag, t$_{lag}$ $\approx$ 32 days. The estimated time lag is bit lower than the observed time lag between optical/UV and X-ray emission, that is $\sim$ 50 days.

An alternative explanation of positive time lag between X-ray and optical-UV is also suggested in \citet{Urry1997}. They proposed that the particles of different energies located at different locations along the jet axis can cause this time lag, high energy particles emitting the X-ray and low energy particles emitting optical-UV.

In this work, we find that the source at different flaring states and at various wavebands are correlated differently with positive and negative time lags and also the non-linear correlation between X-ray and optical-UV suggest that the nature of the source is very complex and hence a detailed temporal study with high quality data with high cadence is needed.

\section{Color-magnitude variation}
Color-magnitude diagram helps to understand the IR-optical-UV emission in blazars. Different types of blazars show different trends such as, FSRQs mostly show "redder-when-brighter" (\citealt{Bonning2012}) behavior and BL Lac show "bluer-when-brighter" (\citealt{Ikejiri2011}) behavior in bright state. Though sometimes same source can show both trend during different bright states.
The optical spectral index can be estimated using the following relation (\citealt{2015A&A...573A..69W})
\begin{equation}
 \alpha_{BR}  = \frac{0.4 (B-R) } {log(\nu_B/\nu_R)}
\end{equation}
Where (B-R) is the color from B and R bands, and the $\nu_B$, $\nu_R$ are the effective frequencies of the corresponding B, and R bands. The factor 0.4 in the numerator comes from the scaling of different bands (\citealt{1998A&A...333..231B}). 

\begin{figure}
    \centering
    \includegraphics[scale=0.5]{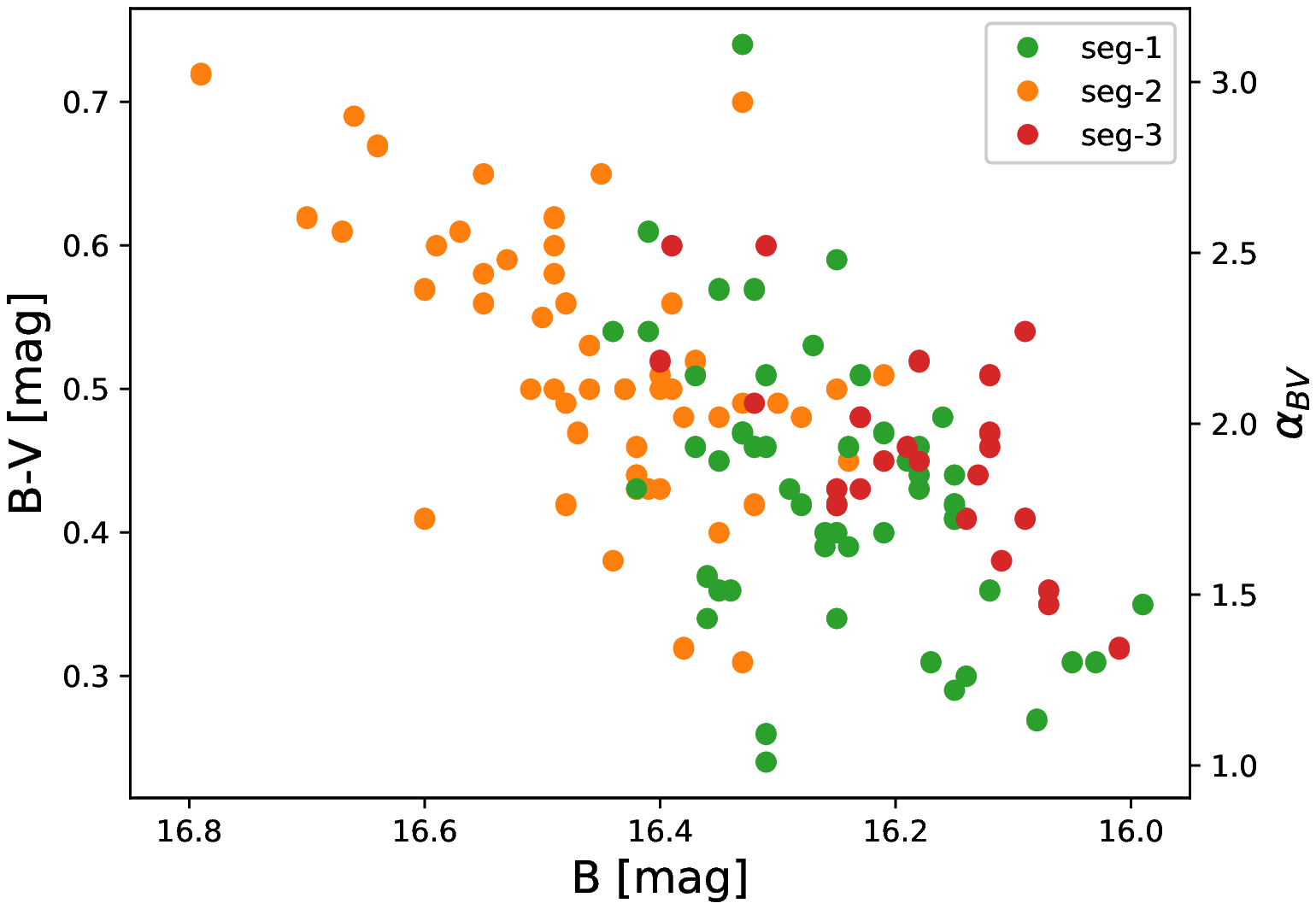}
    \includegraphics[scale=0.5]{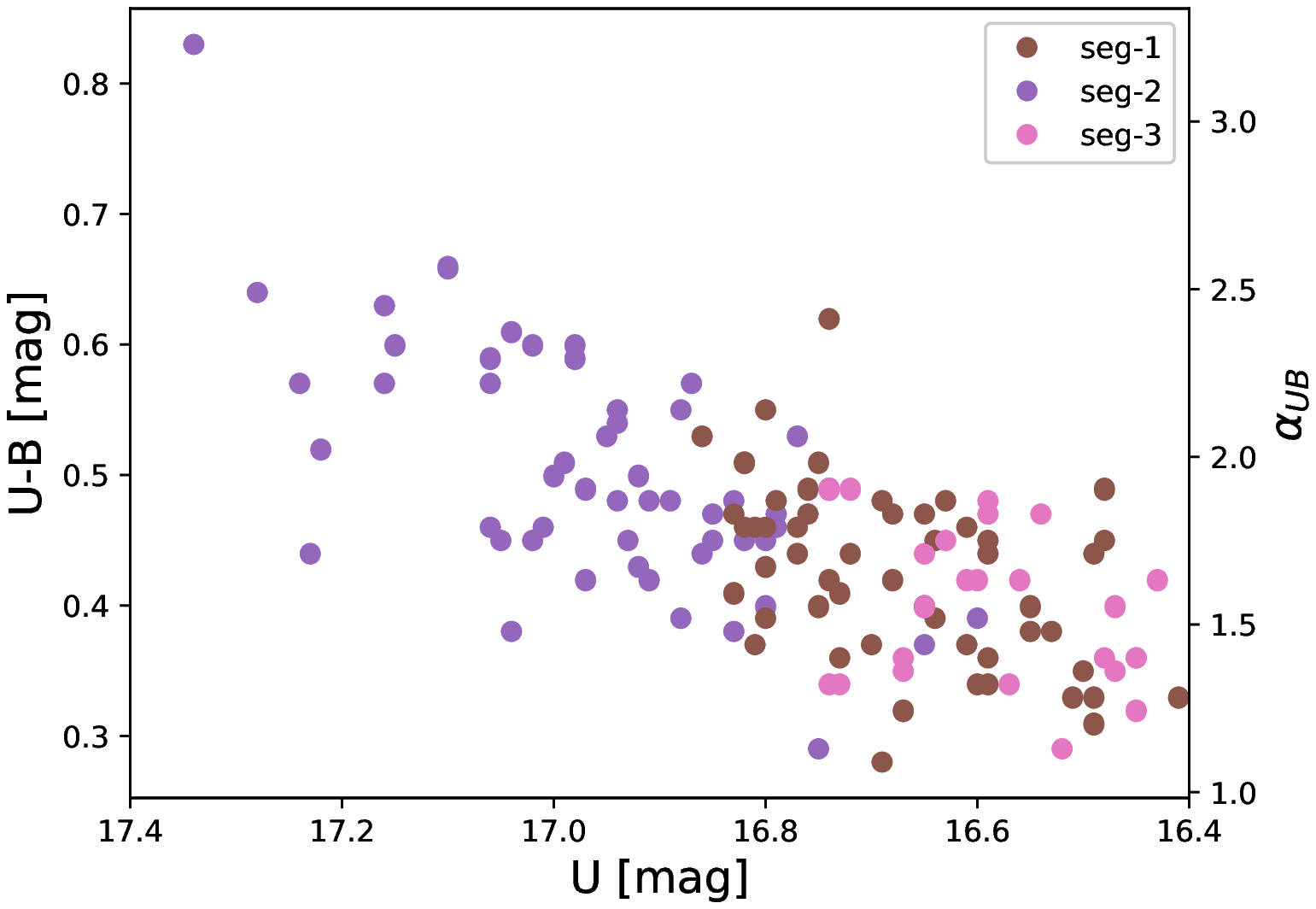}
    \includegraphics[scale=0.5]{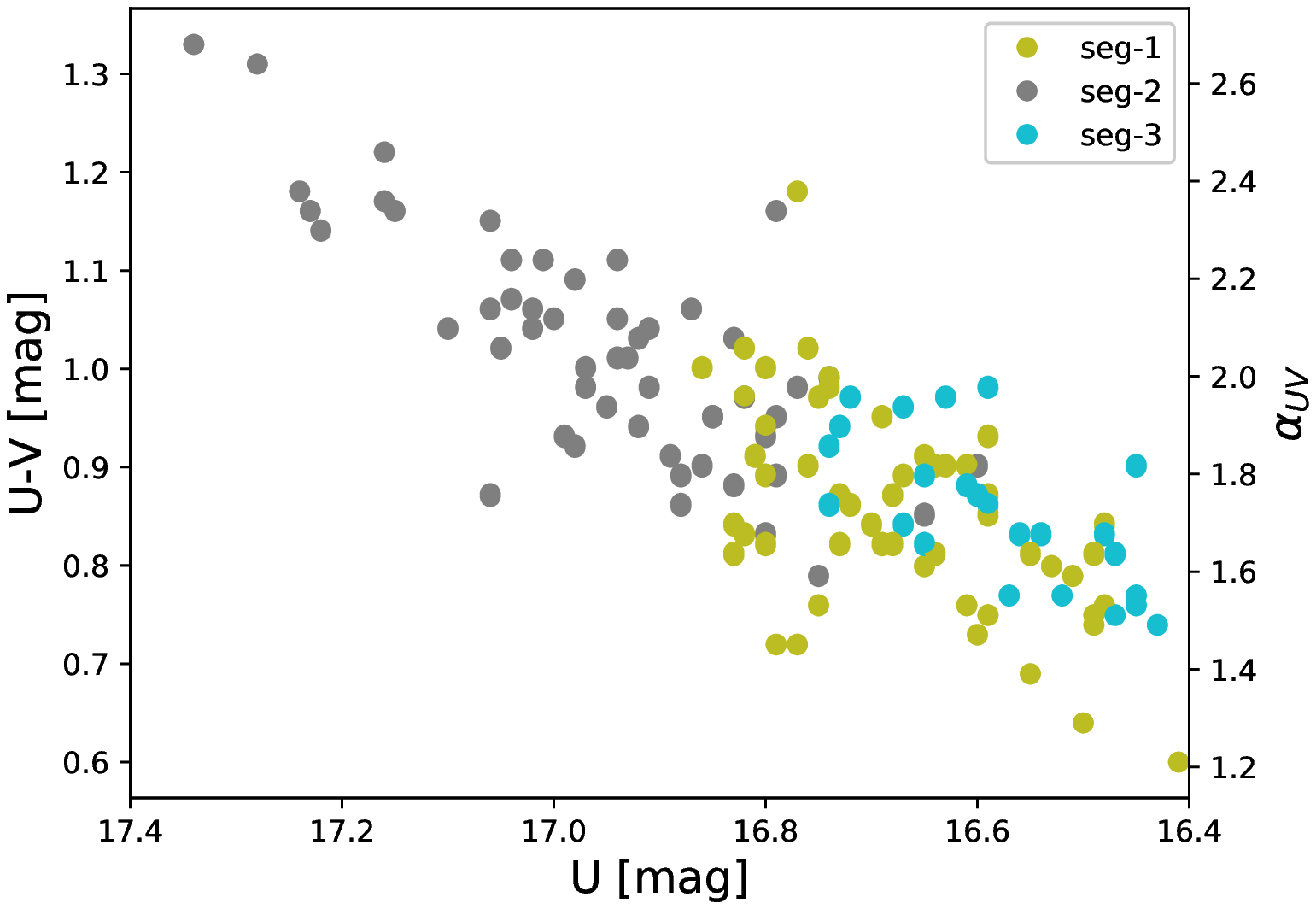}
    \caption{Color-magnitude variation of the various segments 1, 2, and 3 as reported in Figure \ref{fig1}.}
    \label{fig6}
\end{figure}

In Figure \ref{fig6}, we present the color-magnitude (CM) relation with the optical-UV spectral index. To investigate the physical differences between the Segment-1, Segment-2, and Segment-3, their CM diagram is plotted together. As shown in Figure \ref{fig1}, source is bright in optical-UV during Segment-1 and Segment-3, while for Segment-3 there is no bright state in X-ray.
Segment-3 is identified as orphan optical-UV flaring state. The color variation B-V, U-B, and U-V are shown in Figure \ref{fig6} and higher value of color represents the redder behavior and lower value as bluer behavior. It is seen that during the low optical state (segment-2) a "bluer-when-brighter" trend is observed and the color B-V is more scatter than U-B and U-V. A similar trend is observed for Segment-1 and segment-3 also, implying no clear difference between these states (Segment-2 and others). 
An orphan flare is detected in Segment-3, therefore it is expected that the source behaviour in this particular segment would be different than the in other segments, 
but surprisingly there is not much difference. Similar results were also obtained for blazar PKS 0208-512 during the orphan NIR flare and the normal flaring states reported in \citet{Chatterjee2013}.
Study by \citet{Bonning2012}, suggests that most of the FSRQ show "redder-when-brighter" trend during the high state implying a redder jet emission. However, "bluer-when-brighter" trend is very common among the BL Lac type blazars (\citealt{Ikejiri2011}). As it is believed that the disk emission is mostly bluer (bluer than the optical-UV from synchrotron) and the jet emission produced by the Compton scattering are mostly redder. In this case, 
we did not observe any gamma-ray activity suggesting jet emission is constant and the increase in the disk emission can cause the "bluer-when-brighter" trend. To explain the observed "bluer-when-brighter" behavior it could be one of the possibility. 
 In a recent study by \citet{Safna2020} they found a complex nature of color variation where the jet emission for FSRQ and BL Lacs are indistinguishable compared to what \citet{Bonning2012} reported.
In addition, an increase in the accretion disk emission could be accompanied by jet emission produced by the scattering of disk photons. But unfortunately, we did not observe any strong gamma-ray activity during this period. As \citet{Bonning2012} suggests, if the jet is quite brighter then the further brightening due to scattering of high energy population of electrons can also result in a bluer color. This could be true for Segment-1, where the X-ray flare is believed to be produced in the jet and a "bluer-when-brighter" color is observed.

\section{Broadband SED modeling}
Broadband SED modeling is one of the ways to understand the underlying physical mechanisms responsible for broadband emission during flaring as well as low flux states. Simultaneous or quasi-simultaneous coverage of broadband emissions are required to perform such study. With time, the coordination between the Fermi-LAT and the Swift telescope helps to observe a large number of blazars in multibands. 
Many TeV blazars have been detected with Fermi-LAT and few of them have been observed with ground based very high energy gamma-ray telescopes such as MAGIC, H.E.S.S., and VERITAS in TeV energy. The study done in the past using the TeV data prefers mostly the synchrotron self-Compton (SSC) model to explain the high energy peak of the SED. However, in some cases proton-synchrotron has also been used but there is always a problem of overproducing the very high magnetic field in the jet (\citealt{1es1959}). Hadronic modeling with proton-photon and proton-proton interactions are also preferred in few cases where neutrinos are detected or proposed to be detected along with very high energy gamma-ray (\citealt{Bottcher2013}).
The detection of very high energy photons from TeV blazars suggests that the emissions are produced at higher distances from the broad-line region (BLR) in order to avoid the photon-photon absorption in BLR region. Under the \citet{BLANDFORD19871} and \citet{Spitkovsky_2008} model the shock can accelerate the particle to very high energy and are mainly responsible for the high energy emission. However, in case of TeV blazars where emissions are produced at higher distances, the shocks are very weak and an additional model is required to accelerate the charged particles to get high energy emission from them. The preferred model at higher distances is magnetic re-connection (\citealt{Blandford2019}) where small plasmaoids with strong magnetic field can collide together to form a big plasmoid and can act as a region for the acceleration of charged particles (\citealt{Shukla2020}). Sometimes people have also proposed the model of two blobs, where one blob is inside the other blob, and their interaction can accelerate the charged particles responsible for high energy emission {\bf (\citealt{1998A&A...333..452K, MAGIC2020}).}

Here, we have used one-zone SSC modeling to explain the broadband SED for all the three segments. For the SED modeling we have used the specific period from each segment as marked with navy blue color patch in Figure \ref{fig3}. The period chosen for SED modeling during Segment-1 \& 2 is based on high X-ray flux state while in Segment-3, it is chosen for high optical-UV flux state, in order to compare their fitted SED parameters.
During the Segment-1, photons above 300 GeV were observed by MAGIC (\citealt{MAGIC2020}), and it is covered in the period we modeled (MJD 57336-57346). The TeV spectral data points, taken from \citet{MAGIC2020}, are EBL corrected and used in this modeling study. To the best of our knowledge, high energy photons were not detected by MAGIC collaboration for Segment-2 and Segment-3. We also noted very few high energy gamma-ray photons in the Fermi-LAT during the periods marked for SED modeling, and hence many upperlimits are present in the gamma-ray SEDs. The X-ray and optical-UV SEDs are produced as discussed in their analysis section.

\begin{figure*}
    \centering
    \includegraphics[scale=0.35]{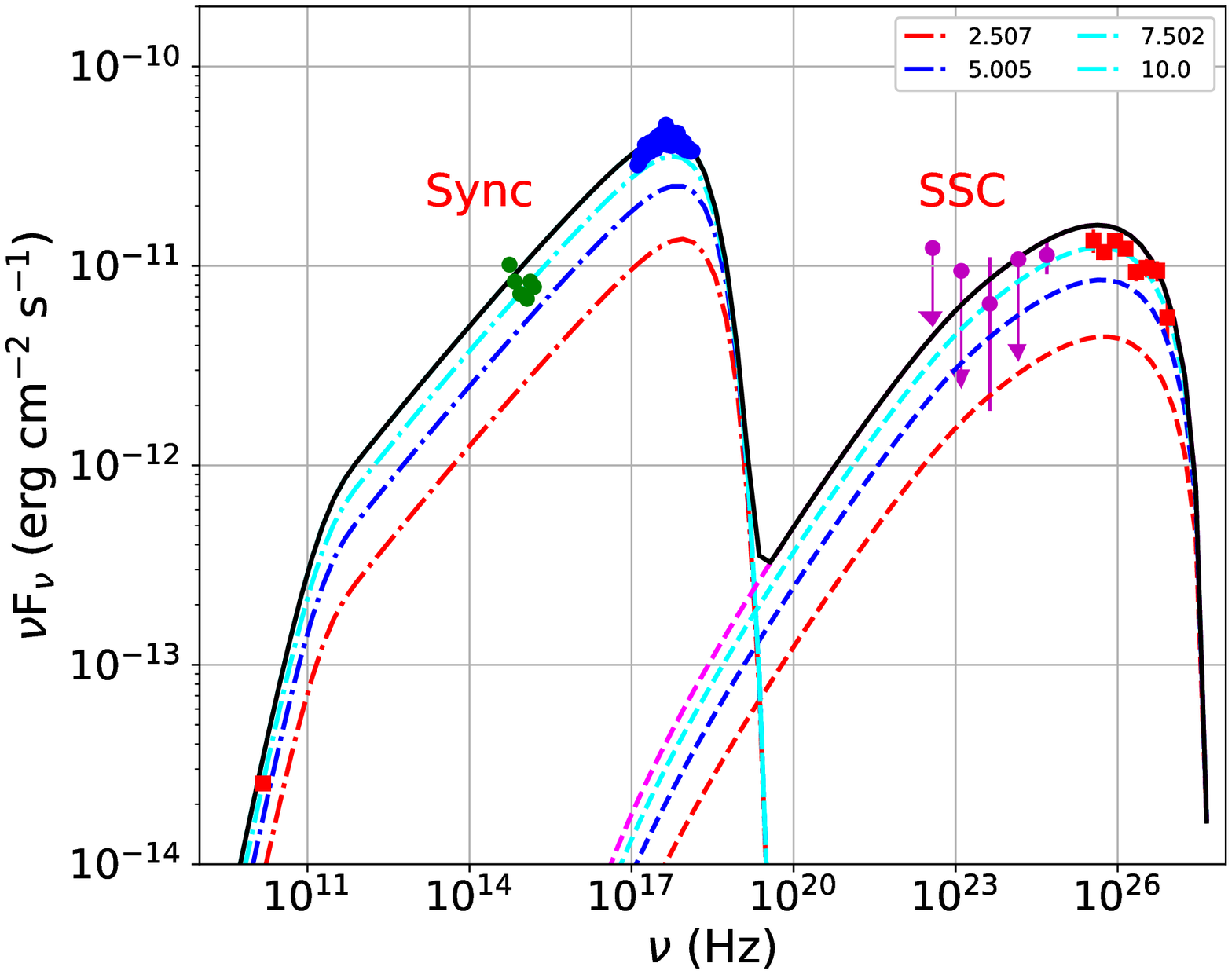}
    \includegraphics[scale=0.35]{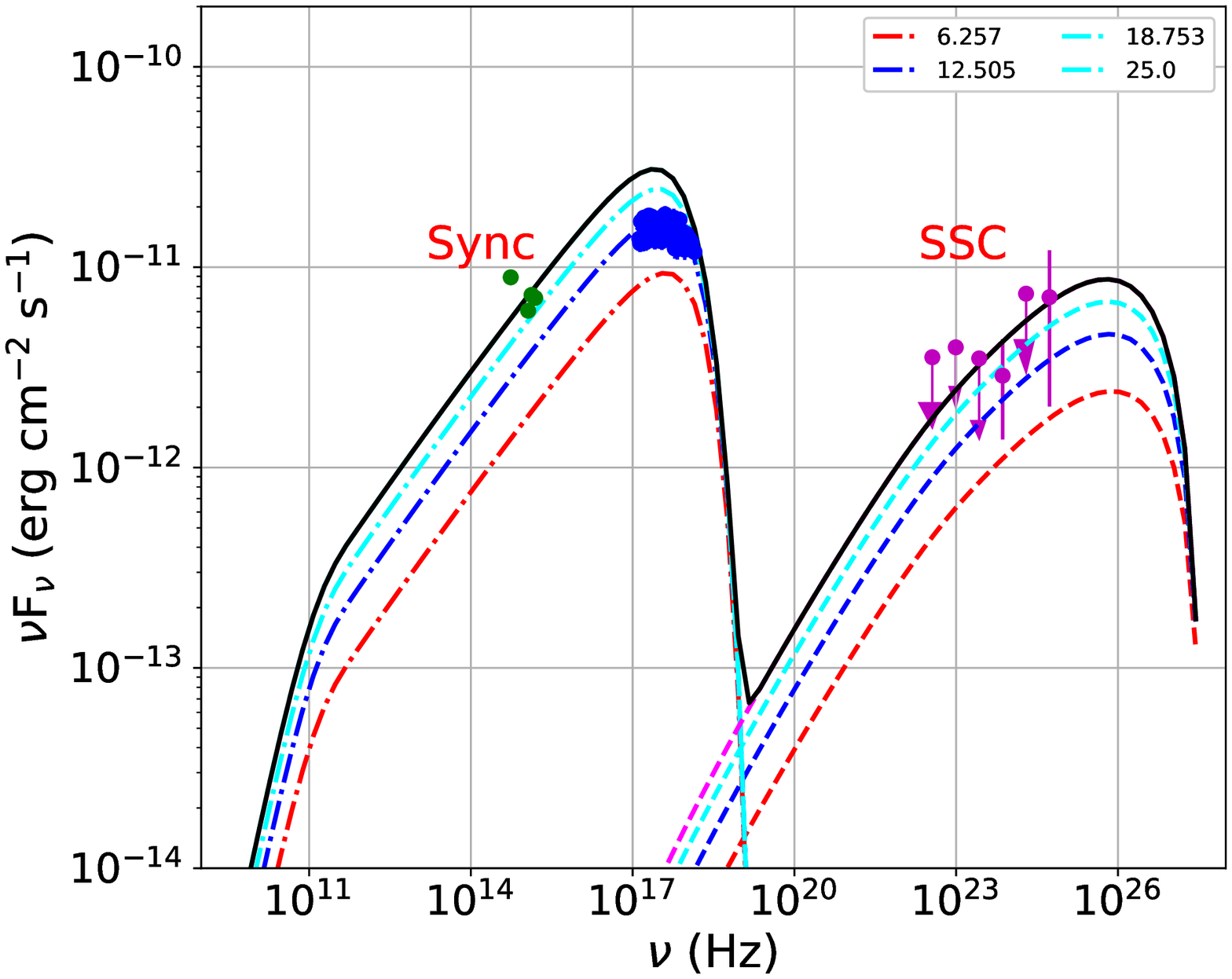}
    \includegraphics[scale=0.35]{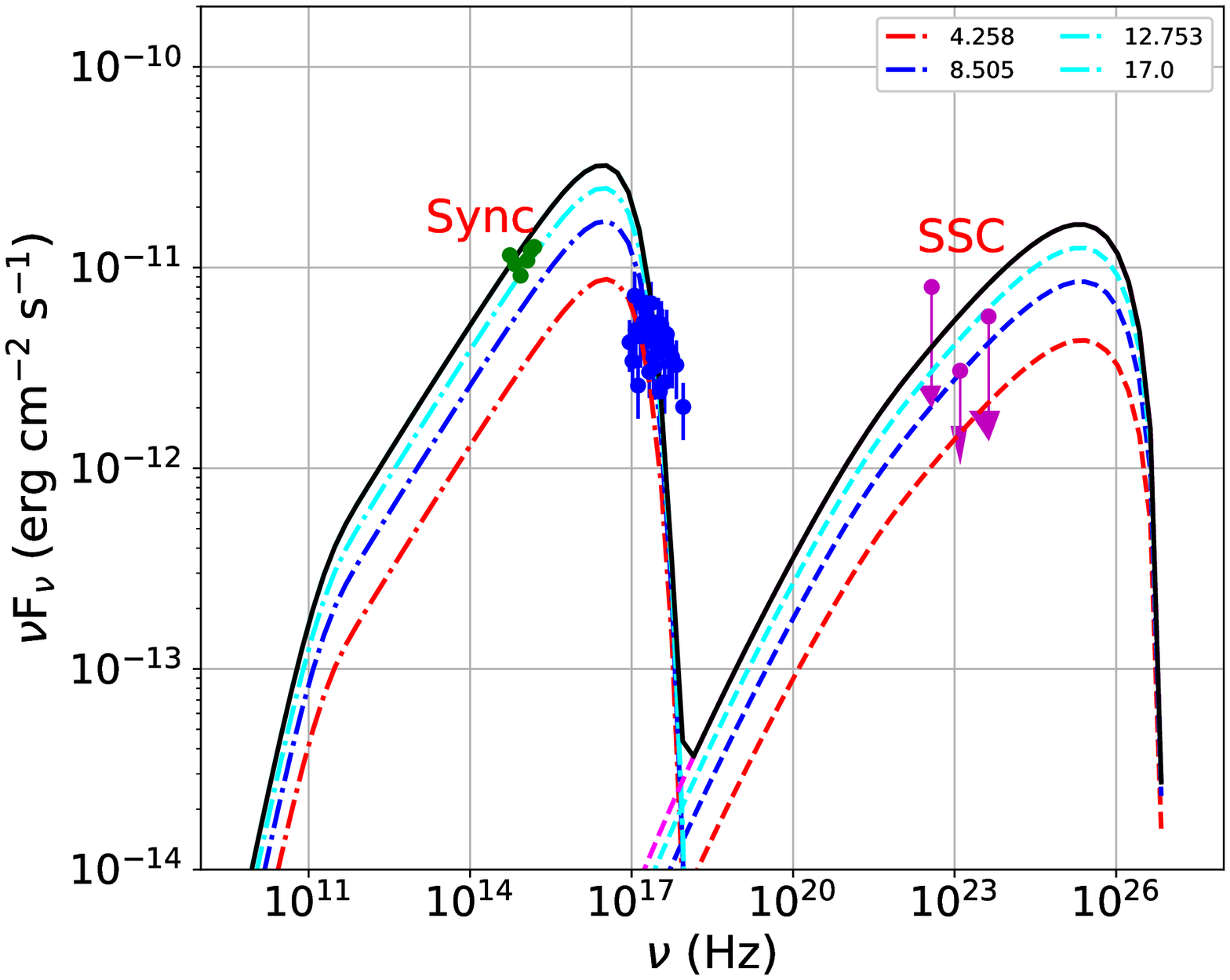}
    \caption{Broadband SED modeling of various states with top panel showing Segment-1(left) and Segment-2(right) and lower panle showing the MWL SED of Segment-3. In Segment-1, red data points represent the MAGIC observations and radio observations from the \citet{MAGIC2020}. The MAGIC data is already corrected for the EBL absorption. Different colors here represent the modeled SED at different time steps (in days) and these times are shown in the legend of each plots. }
    \label{fig7}
\end{figure*}

To model the broadband emission, a publicly available code, GAMERA\footnote{\href{http://libgamera.github.io/GAMERA/docs/documentation.html}{GAMERA}} (\citealt{Hahn2015}), is used. This is a time dependent model, where the charged particle spectrum evolves with time and estimates the photon flux at each given time (for details see  \citealt{Prince_2018, Prince2019a, Prince2019b}). The results of the SED modeling are plotted in Figure \ref{fig7}. Different colors show the SED estimated at various times (in days) and eventually a total sum of the SED is shown with black color. The whole modeling procedure involves many jet parameters such as electrons minimum and maximum energy, electron population distribution spectral index, strength of magnetic field in blob, size of the blob, Lorentz factor and Doppler factor of the blob. The value of Doppler factor ($\delta$) is taken from \citet{MAGIC2020} for the case of single zone, which is, $\delta$=29. If the jet is highly collimated in a small opening angle, then $\theta$ $\sim$ 1/$\Gamma$, and $\delta \sim \Gamma$, where $\Gamma$ is the Lorentz factor.
All the above jet parameters are optimized to achieve the best SED fit to the data and their best fit parameters are presented in Table \ref{tab:sed_param}. 
The size of the emission region is optimized between 0.80$\times$10$^{16}$ cm to 1.16$\times$10$^{16}$ cm for Segment-1, 2, \& 3. A similar size of the emission region is also considered in \citet{MAGIC2020} with value, r=0.7$\times$10$^{16}$ cm. The strength of the magnetic field is found to be in a range 0.02$-$0.03 Gauss, consistent with the broadband modeling done in \citet{MAGIC2020} with single-zone model. The spectral slope of the injected particle distribution is optimized to be, $\alpha$=2.10 for low flux state during Segment-3 to $\alpha$=2.30 for high flux state in Segment-1. The lower energy of the injected electron is constrained by the radio data shown in Segment-1, while for Segment-2 \& 3 it is fixed to the same value, which is, $\gamma_{min}$=150. The maximum energy of the particles changes during various states, and noted to increase with flux states. During Segment-3 (low X-ray flux state), $\gamma_{max}$ = 1.5$\times$10$^5$, and it increased to, $\gamma_{max}$ = 9.3$\times$10$^5$ in Segment-1 (high X-ray flux state), suggesting high energy particles are required to produce the high flux state in X-ray and optical-UV. The power carried by the various components such as electrons, protons, and the magnetic field are consistent with the power estimated in \citet{MAGIC2020} for a single-zone model. The total jet power is well below the Eddington luminosity of the source in all the three states. Following the \citealt{Franceschini2008}, we also noted that the Fermi-LAT spectrum in GeV energy is not affected by the EBL absorption as the source has a very low redshift of 0.055, however, MAGIC data points are corrected for the EBL absorption. 

During the SED modeling, we noted that the location of the synchrotron peak in all the three cases are different. During Segment-3, 
in the low flux state the synchrotron emission peaks at before $\sim$1$\times$10$^{17}$ Hz and the SSC peaks at before $\sim$1$\times$10$^{26}$ Hz. During the Segment-1 and Segment-2, which are the 1st and 2nd highest X-ray flux state, the synchrotron emission peaks at after $\sim$1$\times$10$^{17}$ Hz with SSC peak after the $\sim$1$\times$10$^{26}$ Hz almost at the same place. 
Shift in the synchrotron and the SSC peaks towards higher energy during the high X-ray flux state suggests the involvement of high energy charged particles. The location of synchrotron and SSC peaks during the high states are consistent with the value estimated in \citet{MAGIC2020} and higher than the values reported in \citet{Nilsson2018} (2.2$\times$10$^{16}$ Hz) and \citet{Ajello_2020} (7.5$\times$10$^{15}$ Hz) for this source. As seen in Figure \ref{fig1}, Segment-3 clearly shows an orphan flare in optical-UV with no counterpart in X-ray and $\gamma$-ray. Segment-1 has flaring episode in all the wavebands. Despite having this contrary behavior, the jet parameters derived during these two states are surprisingly quite consistent.

The multi-wavelength campaign of high energy detection was performed in 2013 after VERITAS detection of high energy $\gamma$-ray emission and the broadband SED modeling was done with one-zone SSC mechanism. They derived the size of the emission region of the order of 10$^{17}$ cm with the magnetic field fixed between 0.3-0.6 mGauss. However, the total power derived in the jet is order of 10$^{44}$ erg/s, which is very similar to the value obtained in this study during the high flux state.
MAGIC detection of high energy $\gamma$-ray from 1ES 1727+502 in 2011 along with non-simultaneous archival data collected in optical-UV and X-rays are modeled together with one-zone SSC mechanism by \citet{Aleksic2014}. They obtained the jet parameters such as size of the emission region, R = 7$\times$10$^{15}$cm, magnetic field, B = 0.1 Gauss, and the Doppler factor, $\delta$=15. The size of the emission region is consistent with the value we obtained from the modeling.


\begin{table*}
\centering
\caption{Multiwavelength SED modeling results with the best-fit parameters values. The input injected electron distribution is LogParabola with reference energy 60 MeV. The Doppler factor and the Lorentz factor are fixed at 29.0 considering $\delta$ $\sim$ $\Gamma$. }
 \begin{tabular}{c c c c c}
 \hline \noalign{\smallskip}
Various states& Parameters & Symbols & Values & Period \\
\noalign{\smallskip}  \hline  \noalign{\smallskip}
 Segment-1 & &&& 10 days\\
 & Size of the emitting zone& r & 8.0$\times$10$^{15}$ cm & \\
 & Min Lorentz factor of emitting electrons & $\gamma_{min}$& 150.0 &\\
 & Max Lorentz factor of emitting electrons & $\gamma_{max}$& 9.3$\times$10$^{5}$ &\\
 & Input injected electron spectrum (LP) & $\alpha$ & 2.30 & \\
 & Curvature parameter of the PL spectrum & $\beta$& 0.02 & \\
 & Magnetic field in emitting zone & B & 0.03 G & \\
 & Jet power in electrons & P$_{j,e}$ & 2.36$\times$10$^{44}$ erg/s & \\
 & Jet power in magnetic field & P$_{j,B}$ & 1.81$\times$10$^{41}$ erg/s & \\
 & Jet power in protons & P$_{j,P}$ &4.19$\times$10$^{43}$ erg/s& \\
 & Total jet power & P$_{jet}$ & 2.78$\times$10$^{44}$ erg/s& \\
 \noalign{\smallskip} \hline   \noalign{\smallskip}
Segment-2 & &&& 25 days \\
& Size of the emitting zone& r & 1.16$\times$10$^{16}$ cm & \\
 & Min Lorentz factor of emitting electrons & $\gamma_{min}$& 150.0 &\\
 & Max Lorentz factor of emitting electrons & $\gamma_{max}$& 7.3$\times$10$^{5}$ &\\
 & Input injected electron spectrum (LP) & $\alpha$ & 2.20 & \\
 & Curvature parameter of the PL spectrum & $\beta$& 0.02 & \\
 & Magnetic field in emitting zone & B & 0.02 G & \\
 & Jet power in electrons & P$_{j,e}$ & 6.96$\times$10$^{43}$ erg/s & \\
 & Jet power in magnetic field & P$_{j,B}$ & 1.53$\times$10$^{41}$ erg/s & \\
 & Jet power in protons & P$_{j,P}$ & 1.06$\times$10$^{43}$ erg/s& \\
 & Total jet power & P$_{jet}$ & 8.01$\times$10$^{43}$ erg/s& \\
 \noalign{\smallskip}  \hline  \noalign{\smallskip}
 Segment-3 & &&& 17 days \\
 & Size of the emitting zone& r & 8.0$\times$10$^{15}$ cm & \\
 & Min Lorentz factor of emitting electrons & $\gamma_{min}$& 150.0 &\\
 & Max Lorentz factor of emitting electrons & $\gamma_{max}$& 1.5$\times$10$^{5}$ &\\
 & Input injected electron spectrum (LP) & $\alpha$ & 2.10 & \\
 & Curvature parameter of the PL spectrum & $\beta$& 0.02 & \\
 & Magnetic field in emitting zone & B & 0.03 G & \\
 & Jet power in electrons & P$_{j,e}$ & 1.05$\times$10$^{44}$ erg/s & \\
 & Jet power in magnetic field & P$_{j,B}$ & 1.81$\times$10$^{41}$ erg/s & \\
 & Jet power in protons & P$_{j,P}$ & 1.41$\times$10$^{43}$ erg/s& \\
 & Total jet power & P$_{jet}$ & 1.19$\times$10$^{44}$ erg/s& \\
\noalign{\smallskip} \hline  \noalign{\smallskip}
 \end{tabular}
 \label{tab:sed_param}

\end{table*}

\section{Summary and Discussions}
In this work, we present the broadband temporal and spectral study of TeV blazar 1ES 1727+502 
using 8-years of long data from 2014 to 2021.
A strong long-term X-ray flaring activity is observed at the end of 2014 and beginning of 2015 followed by a bright state in optical and UV. A flare is observed in $\gamma$-ray also during the same period but the exact time of the $\gamma$-ray flare is missed in optical-UV and X-ray observations. After that, long-term X-ray and optical observations show small variation in their fluxes while in $\gamma$-ray no activity is recorded simultaneously.
The long-term light curves are divided into three parts namely, Segment-1, 2, and 3. Segment-3 is recognised as a period of the orphan optical-UV flares where X-ray and $\gamma$-ray fluxes are quite moderate.
A significant amount of flux variability is observed in $\gamma$-ray and X-ray followed by optical and UV in Segment-1. During Segment-2 and Segment-3 a very low flux variability is seen, whereas the source is more variable in X-ray and optical-UV than in $\gamma$-ray. The shape of the fractional variability resembles the broadband SED shape and it is believed that the less variable energy bands relates to the rising part of the SED humps and more variable energy bands relates tot he decaying part of the SED humps (\citealt{2015A&A...576A.126A, 2015A&A...578A..22A}). 
In Segment-3, due to the orphan optical-UV flare a significant variability is recorded in these bands compared to X-ray and $\gamma$-ray. The variability study reveals the involvement of range of variability time across the wavebands starting from fast flux variability of days scale to the longer time scale variability of month scale. 
The spectral variation of 
X-ray and $\gamma$-ray with respect to corresponding flux values suggest an opposite trend. In X-ray, it shows a hint of "harder-when-brighter" trend, but not clear, and in $\gamma$-ray it shows "softer-when-brighter" trend quantified by Pearson's correlation coefficient. The spectral indices derived for optical wavebands from the color-magnitude diagram also suggest the "harder-when-brighter" trend. Similar trend seen in X-ray and optical emissions suggest that both the emissions are produced by the similar process under the same circumstances, in this case it is synchrotron mechanism. The CM variations among the combinations of optical bands show a "bluer-when-brighter" chromatism among all the three segments. The broadband correlation is done for all the three segments separately. Segment-1 light curve is divided into two parts and the flux in these two parts are separately correlated with optical-UV. In part-1 optical-UV emission leads the X-ray emission by 50 days and in part-2, X-ray leads the optical-UV by 25 and 50 days. During part-2 light curve of the Segment-3, we observed the optical-UV emission leads the X-ray band emission by similar time interval, $\sim$ 50 days.
Under the leptonic scenario, the optical-UV and the X-rays are produced at the same time as also have been seen in many blazars. The time delay observed between optical-UV could be explained by the two scenarios, first: either there are two emission regions separated by a distance corresponding to 50 days time delay and emit simultaneously. 
Secondly, we can assume that the emissions are produced close to the base of the jet where the optical depth of the region is very high and as a result the optical emission can be suppressed. While the emission region travels a bit farther from the base of the jet, the optical emission is observed and hence a time delay is seen.
Other possibility is, if the emission region is extended to a larger size then the synchrotron emission produced by the charge particle motion around the magnetic field can results in emissions on multiple time-scales.
The part of the light curves which do not have sufficient number of data points for cross-correlation, the Pearson's correlation is estimated. The flux-flux correlation results suggest that optical-UV are highly correlated, however, X-ray and UV correlation show a complex nature. The flux-flux correlation is very common in AGN where the emission is dominated by the central part of the source \citealt{2020ApJ...896....1C}.
 The observed time lags of different time-scale among various wavebands implies that the nature of the emission is very complex. To better understand the source and its flaring activity good cadence data is required in optical-UV and X-ray.
To explore the physical nature of the broadband emission, a broadband SED modeling is performed considering all the possible physical mechanisms. Since it is a BL Lac source a well accepted one-zone SSC model is used and the data is well fitted with this model.
The jet parameters derived during the various states are consistent with each other and they also agree with the earlier studies where SED modeling was done for this source (\citet{Aleksic2014}, \citet{Archambault2015}, \citet{MAGIC2020}). The detection of TeV $\gamma$-ray photons as reported in earlier studies suggest that this source could also be a possible source of astrophysical neutrinos under the hadronic scenario. Many more, broadband SED modeling and detailed study of TeV photons are
required to acquire the proper inference from these sources. 

\section{Conclusions}

We conclude the present study as follows, 
\begin{itemize}
    \item The total light curve between 2014-2021 is divided into three parts such as Segment-1, 2, \& 3.
    The brightest X-ray and optical-UV flare is observed during 2014-2015 with highest variability and the fastest variability time. The pattern of the fractional variability resembles the structure of the broadband SED. 
    \item Segment-3 witnessed an orphan flare in optical-UV with no association in X-ray and $\gamma$-rays. A very rare phenomena which question our understanding of blazar and the physical processes behind its broadband emission.
    \item The optical-UV emission shows a clear "harder-when-brighter" trend and the $\gamma$-ray emission 
    shows the opposite trend, e.g. "softer-when-brighter". However, in X-rays trend is not clear. 
    \item The color-magnitude variation reveals the "bluer-when-brighter" chromatism suggesting the emission is dominated by the disk rather than the jet (jet has a constant emission). However, the boradband SED modeling consider the jet emission and provides a good fit to the data.
    \item One part of the light curves shows the X-ray leading the optical-UV emissions by $\sim$ 50 days and the other part shows the opposite behavior with optical-UV leading the X-rays by similar amount of time delay.
    \item The flux correlation between optical, UV and the X-ray reveals that the optical \& UV are highly correlated. However, emissions in X-ray and UV show a 
    complex correlation (not following a straight line in flux-flux plot.)
    \item The broadband SED modeling suggests that the jet parameters during different flux states appear to be similar. 
    \item During the high flux states high energy of charged particles are required to explain the high energy part of the spectrum.
\end{itemize}

\section*{Acknowledgements}
We thank the referee for his/her helpful comments and suggestions during review process.
RP thanks Prof. Ranjeev Misra from IUCCA for the helpful scientific discussions.
The project is partially supported by the Polish Funding Agency National Science Centre, project 2017/26/A/ST9/00756 (MAESTRO 9), and MNiSW grant DIR/WK/2018/12. PM acknowledges the generous support of the Stanislaw Ulam fellowship (PPN/ULM/2019/1/00096/A/00001) by Polish National Agency for Academic Exchange (NAWA). 

\section*{Data Availability}

The X-ray and optical-UV data and related softwares used in this research are available at NASA’s HEASARC webpages, with the links given in the manuscript. The gamma-ray data from Fermi-LAT is also available on Fermi-webpage provided in manuscript. The link for time dependent SED code to model the broadband SED is also provided in manuscript.



\bibliographystyle{mnras}
\bibliography{example} 

\begin{thebibliography}{}
\makeatletter
\relax
\def\mn@urlcharsother{\let\do\@makeother \do\$\do\&\do\#\do\^\do\_\do\%\do\~}
\def\mn@doi{\begingroup\mn@urlcharsother \@ifnextchar [ {\mn@doi@}
  {\mn@doi@[]}}
\def\mn@doi@[#1]#2{\def\@tempa{#1}\ifx\@tempa\@empty \href
  {http://dx.doi.org/#2} {doi:#2}\else \href {http://dx.doi.org/#2} {#1}\fi
  \endgroup}
\def\mn@eprint#1#2{\mn@eprint@#1:#2::\@nil}
\def\mn@eprint@arXiv#1{\href {http://arxiv.org/abs/#1} {{\tt arXiv:#1}}}
\def\mn@eprint@dblp#1{\href {http://dblp.uni-trier.de/rec/bibtex/#1.xml}
  {dblp:#1}}
\def\mn@eprint@#1:#2:#3:#4\@nil{\def\@tempa {#1}\def\@tempb {#2}\def\@tempc
  {#3}\ifx \@tempc \@empty \let \@tempc \@tempb \let \@tempb \@tempa \fi \ifx
  \@tempb \@empty \def\@tempb {arXiv}\fi \@ifundefined
  {mn@eprint@\@tempb}{\@tempb:\@tempc}{\expandafter \expandafter \csname
  mn@eprint@\@tempb\endcsname \expandafter{\@tempc}}}

\bibitem[\protect\citeauthoryear{Abdo et~al.,}{Abdo et~al.}{2010}]{Abdo_2010}
Abdo A.~A.,  et~al., 2010, \mn@doi [The Astrophysical Journal]
  {10.1088/0004-637x/716/1/30}, 716, 30

\bibitem[\protect\citeauthoryear{{Abdo} et~al.,}{{Abdo}
  et~al.}{2011}]{2011ApJ...736..131A}
{Abdo} A.~A.,  et~al., 2011, \mn@doi [\apj] {10.1088/0004-637X/736/2/131},
  \href {https://ui.adsabs.harvard.edu/abs/2011ApJ...736..131A} {736, 131}

\bibitem[\protect\citeauthoryear{Abdollahi et~al.,}{Abdollahi
  et~al.}{2020}]{Abdollahi_2020}
Abdollahi S.,  et~al., 2020, \mn@doi [The Astrophysical Journal Supplement
  Series] {10.3847/1538-4365/ab6bcb}, 247, 33

\bibitem[\protect\citeauthoryear{{Ackermann} et~al.,}{{Ackermann}
  et~al.}{2016}]{Ackermann2016}
{Ackermann} M.,  et~al., 2016, \mn@doi [\apjl] {10.3847/2041-8205/824/2/L20},
  \href {https://ui.adsabs.harvard.edu/abs/2016ApJ...824L..20A} {824, L20}

\bibitem[\protect\citeauthoryear{Ajello et~al.,}{Ajello
  et~al.}{2020}]{Ajello_2020}
Ajello M.,  et~al., 2020, \mn@doi [The Astrophysical Journal]
  {10.3847/1538-4357/ab791e}, 892, 105

\bibitem[\protect\citeauthoryear{Aleksi{\'{c}} et~al.,}{Aleksi{\'{c}}
  et~al.}{2011}]{Aleksi_2011}
Aleksi{\'{c}} J.,  et~al., 2011, \mn@doi [The Astrophysical Journal]
  {10.1088/0004-637x/729/2/115}, 729, 115

\bibitem[\protect\citeauthoryear{{Aleksi{\'c}} et~al.,}{{Aleksi{\'c}}
  et~al.}{2014}]{Aleksic2014}
{Aleksi{\'c}} J.,  et~al., 2014, \mn@doi [\aap] {10.1051/0004-6361/201321360},
  \href {https://ui.adsabs.harvard.edu/abs/2014A&A...563A..90A} {563, A90}

\bibitem[\protect\citeauthoryear{{Aleksi{\'c}} et~al.,}{{Aleksi{\'c}}
  et~al.}{2015a}]{2015A&A...576A.126A}
{Aleksi{\'c}} J.,  et~al., 2015a, \mn@doi [\aap] {10.1051/0004-6361/201424216},
  \href {https://ui.adsabs.harvard.edu/abs/2015A&A...576A.126A} {576, A126}

\bibitem[\protect\citeauthoryear{{Aleksi{\'c}} et~al.,}{{Aleksi{\'c}}
  et~al.}{2015b}]{2015A&A...578A..22A}
{Aleksi{\'c}} J.,  et~al., 2015b, \mn@doi [\aap] {10.1051/0004-6361/201424811},
  \href {https://ui.adsabs.harvard.edu/abs/2015A&A...578A..22A} {578, A22}

\bibitem[\protect\citeauthoryear{{Alexander}}{{Alexander}}{2013}]{2013arXiv1302.1508A}
{Alexander} T.,  2013, arXiv e-prints, \href
  {https://ui.adsabs.harvard.edu/abs/2013arXiv1302.1508A} {p. arXiv:1302.1508}

\bibitem[\protect\citeauthoryear{{Archambault} et~al.,}{{Archambault}
  et~al.}{2015}]{Archambault2015}
{Archambault} S.,  et~al., 2015, \mn@doi [\apj] {10.1088/0004-637X/808/2/110},
  \href {https://ui.adsabs.harvard.edu/abs/2015ApJ...808..110A} {808, 110}

\bibitem[\protect\citeauthoryear{{Atwood} et~al.,}{{Atwood}
  et~al.}{2009}]{Atwood2009}
{Atwood} W.~B.,  et~al., 2009, \mn@doi [\apj] {10.1088/0004-637X/697/2/1071},
  \href {https://ui.adsabs.harvard.edu/abs/2009ApJ...697.1071A} {697, 1071}

\bibitem[\protect\citeauthoryear{{Bai} \& {Lee}}{{Bai} \&
  {Lee}}{2003}]{Bai2003}
{Bai} J.~M.,  {Lee} M.~G.,  2003, \mn@doi [\apjl] {10.1086/374304}, \href
  {https://ui.adsabs.harvard.edu/abs/2003ApJ...585L.113B} {585, L113}

\bibitem[\protect\citeauthoryear{{Balokovi{\'c}} et~al.,}{{Balokovi{\'c}}
  et~al.}{2016}]{2016ApJ...819..156B}
{Balokovi{\'c}} M.,  et~al., 2016, \mn@doi [\apj]
  {10.3847/0004-637X/819/2/156}, \href
  {https://ui.adsabs.harvard.edu/abs/2016ApJ...819..156B} {819, 156}

\bibitem[\protect\citeauthoryear{{Begelman} \& {Sikora}}{{Begelman} \&
  {Sikora}}{1987}]{Begelman1987}
{Begelman} M.~C.,  {Sikora} M.,  1987, \mn@doi [\apj] {10.1086/165760}, \href
  {https://ui.adsabs.harvard.edu/abs/1987ApJ...322..650B} {322, 650}

\bibitem[\protect\citeauthoryear{{Bessell}, {Castelli}  \& {Plez}}{{Bessell}
  et~al.}{1998}]{1998A&A...333..231B}
{Bessell} M.~S.,  {Castelli} F.,   {Plez} B.,  1998, \aap, \href
  {https://ui.adsabs.harvard.edu/abs/1998A&A...333..231B} {333, 231}

\bibitem[\protect\citeauthoryear{Blandford \& Eichler}{Blandford \&
  Eichler}{1987}]{BLANDFORD19871}
Blandford R.,  Eichler D.,  1987, \mn@doi [Physics Reports]
  {https://doi.org/10.1016/0370-1573(87)90134-7}, 154, 1

\bibitem[\protect\citeauthoryear{{Blandford} \& {K{\"o}nigl}}{{Blandford} \&
  {K{\"o}nigl}}{1979}]{Blandford1979}
{Blandford} R.~D.,  {K{\"o}nigl} A.,  1979, \mn@doi [\apj] {10.1086/157262},
  \href {https://ui.adsabs.harvard.edu/abs/1979ApJ...232...34B} {232, 34}

\bibitem[\protect\citeauthoryear{{Blandford} \& {Payne}}{{Blandford} \&
  {Payne}}{1982}]{Blandford1982}
{Blandford} R.~D.,  {Payne} D.~G.,  1982, \mn@doi [\mnras]
  {10.1093/mnras/199.4.883}, \href
  {https://ui.adsabs.harvard.edu/abs/1982MNRAS.199..883B} {199, 883}

\bibitem[\protect\citeauthoryear{Blandford, Meier  \& Readhead}{Blandford
  et~al.}{2019}]{Blandford2019}
Blandford R.,  Meier D.,   Readhead A.,  2019, \mn@doi [Annual Review of
  Astronomy and Astrophysics] {10.1146/annurev-astro-081817-051948}, 57, 467

\bibitem[\protect\citeauthoryear{B{\l}a{\.{z}}ejowski, Sikora, Moderski  \&
  Madejski}{B{\l}a{\.{z}}ejowski et~al.}{2000}]{Baejowski2000}
B{\l}a{\.{z}}ejowski M.,  Sikora M.,  Moderski R.,   Madejski G.~M.,  2000,
  \mn@doi [The Astrophysical Journal] {10.1086/317791}, 545, 107

\bibitem[\protect\citeauthoryear{{Boettcher}, {Mause}  \&
  {Schlickeiser}}{{Boettcher} et~al.}{1997}]{Boettcher1997}
{Boettcher} M.,  {Mause} H.,   {Schlickeiser} R.,  1997, \aap, \href
  {https://ui.adsabs.harvard.edu/abs/1997A&A...324..395B} {324, 395}

\bibitem[\protect\citeauthoryear{{Bonning} et~al.,}{{Bonning}
  et~al.}{2012}]{Bonning2012}
{Bonning} E.,  et~al., 2012, \mn@doi [\apj] {10.1088/0004-637X/756/1/13}, \href
  {https://ui.adsabs.harvard.edu/abs/2012ApJ...756...13B} {756, 13}

\bibitem[\protect\citeauthoryear{{Boula} \& {Mastichiadis}}{{Boula} \&
  {Mastichiadis}}{2022}]{2022A&A...657A..20B}
{Boula} S.,  {Mastichiadis} A.,  2022, \mn@doi [\aap]
  {10.1051/0004-6361/202142126}, \href
  {https://ui.adsabs.harvard.edu/abs/2022A&A...657A..20B} {657, A20}

\bibitem[\protect\citeauthoryear{Breeveld, Landsman, Holland, Roming, Kuin  \&
  Page}{Breeveld et~al.}{2011}]{Breeveld_2011}
Breeveld A.~A.,  Landsman W.,  Holland S.~T.,  Roming P.,  Kuin N. P.~M.,
  Page M.~J.,  2011, \mn@doi [AIP Conference Proceedings] {10.1063/1.3621807},
  1358, 373

\bibitem[\protect\citeauthoryear{{Brown}}{{Brown}}{2013}]{Brown2013}
{Brown} A.~M.,  2013, \mn@doi [\mnras] {10.1093/mnras/stt218}, \href
  {https://ui.adsabs.harvard.edu/abs/2013MNRAS.431..824B} {431, 824}

\bibitem[\protect\citeauthoryear{Böttcher, Reimer, Sweeney  \&
  Prakash}{Böttcher et~al.}{2013}]{Bottcher2013}
Böttcher M.,  Reimer A.,  Sweeney K.,   Prakash A.,  2013, \mn@doi [The
  Astrophysical Journal] {10.1088/0004-637x/768/1/54}, 768, 54

\bibitem[\protect\citeauthoryear{{Cackett} et~al.,}{{Cackett}
  et~al.}{2020}]{2020ApJ...896....1C}
{Cackett} E.~M.,  et~al., 2020, \mn@doi [\apj] {10.3847/1538-4357/ab91b5},
  \href {https://ui.adsabs.harvard.edu/abs/2020ApJ...896....1C} {896, 1}

\bibitem[\protect\citeauthoryear{{Cerruti}, {Benbow}, {Chen}, {Dumm}, {Fortson}
   \& {Shahinyan}}{{Cerruti} et~al.}{2017}]{Cerruti2017}
{Cerruti} M.,  {Benbow} W.,  {Chen} X.,  {Dumm} J.~P.,  {Fortson} L.~F.,
  {Shahinyan} K.,  2017, \mn@doi [\aap] {10.1051/0004-6361/201730799}, \href
  {https://ui.adsabs.harvard.edu/abs/2017A&A...606A..68C} {606, A68}

\bibitem[\protect\citeauthoryear{{Chatterjee} et~al.,}{{Chatterjee}
  et~al.}{2013}]{Chatterjee2013}
{Chatterjee} R.,  et~al., 2013, \mn@doi [\apjl] {10.1088/2041-8205/763/1/L11},
  \href {https://ui.adsabs.harvard.edu/abs/2013ApJ...763L..11C} {763, L11}

\bibitem[\protect\citeauthoryear{{Chatterjee} et~al.,}{{Chatterjee}
  et~al.}{2021a}]{Chatterjee2021}
{Chatterjee} R.,  et~al., 2021a, arXiv e-prints, \href
  {https://ui.adsabs.harvard.edu/abs/2021arXiv210200919C} {p. arXiv:2102.00919}

\bibitem[\protect\citeauthoryear{{Chatterjee} et~al.,}{{Chatterjee}
  et~al.}{2021b}]{Ritaban2021}
{Chatterjee} R.,  et~al., 2021b, arXiv e-prints, \href
  {https://ui.adsabs.harvard.edu/abs/2021arXiv210200919C} {p. arXiv:2102.00919}

\bibitem[\protect\citeauthoryear{{Das}, {Prince}  \& {Gupta}}{{Das}
  et~al.}{2021}]{Avik2021}
{Das} A.~K.,  {Prince} R.,   {Gupta} N.,  2021, arXiv e-prints, \href
  {https://ui.adsabs.harvard.edu/abs/2021arXiv210710555D} {p. arXiv:2107.10555}

\bibitem[\protect\citeauthoryear{{Dermer} \& {Schlickeiser}}{{Dermer} \&
  {Schlickeiser}}{1993}]{Dermer1993}
{Dermer} C.~D.,  {Schlickeiser} R.,  1993, \mn@doi [\apj] {10.1086/173251},
  \href {https://ui.adsabs.harvard.edu/abs/1993ApJ...416..458D} {416, 458}

\bibitem[\protect\citeauthoryear{Diltz \& Böttcher}{Diltz \&
  Böttcher}{2016}]{Diltz_2016}
Diltz C.,  Böttcher M.,  2016, \mn@doi [The Astrophysical Journal]
  {10.3847/0004-637x/826/1/54}, 826, 54

\bibitem[\protect\citeauthoryear{{Edelson} \& {Krolik}}{{Edelson} \&
  {Krolik}}{1988}]{Edelson1988}
{Edelson} R.~A.,  {Krolik} J.~H.,  1988, \mn@doi [\apj] {10.1086/166773}, \href
  {https://ui.adsabs.harvard.edu/abs/1988ApJ...333..646E} {333, 646}

\bibitem[\protect\citeauthoryear{{Franceschini}, {Rodighiero}  \&
  {Vaccari}}{{Franceschini} et~al.}{2008}]{Franceschini2008}
{Franceschini} A.,  {Rodighiero} G.,   {Vaccari} M.,  2008, \mn@doi [\aap]
  {10.1051/0004-6361:200809691}, \href
  {https://ui.adsabs.harvard.edu/abs/2008A&A...487..837F} {487, 837}

\bibitem[\protect\citeauthoryear{Ghisellini \& Madau}{Ghisellini \&
  Madau}{1996}]{Ghisellini1996}
Ghisellini G.,  Madau P.,  1996, \mn@doi [Monthly Notices of the Royal
  Astronomical Society] {10.1093/mnras/280.1.67}, 280, 67

\bibitem[\protect\citeauthoryear{{Ghisellini} \& {Maraschi}}{{Ghisellini} \&
  {Maraschi}}{1989}]{Ghisellini1989}
{Ghisellini} G.,  {Maraschi} L.,  1989, \mn@doi [\apj] {10.1086/167383}, \href
  {https://ui.adsabs.harvard.edu/abs/1989ApJ...340..181G} {340, 181}

\bibitem[\protect\citeauthoryear{Ghisellini \& Tavecchio}{Ghisellini \&
  Tavecchio}{2008}]{Ghisellini2008}
Ghisellini G.,  Tavecchio F.,  2008, \mn@doi [Monthly Notices of the Royal
  Astronomical Society] {10.1111/j.1365-2966.2008.13360.x}, 387, 1669

\bibitem[\protect\citeauthoryear{{Giommi} et~al.,}{{Giommi}
  et~al.}{2006}]{Giommi2006}
{Giommi} P.,  et~al., 2006, \mn@doi [\aap] {10.1051/0004-6361:20064874}, \href
  {https://ui.adsabs.harvard.edu/abs/2006A&26A...456..911G} {456, 911}

\bibitem[\protect\citeauthoryear{{Hahn}}{{Hahn}}{2015}]{Hahn2015}
{Hahn} J.,  2015, in 34th International Cosmic Ray Conference (ICRC2015).
  p.~917

\bibitem[\protect\citeauthoryear{{Heidt} \& {Wagner}}{{Heidt} \&
  {Wagner}}{1996}]{Heidt1996}
{Heidt} J.,  {Wagner} S.~J.,  1996, \aap, \href
  {https://ui.adsabs.harvard.edu/abs/1996A&A...305...42H} {305, 42}

\bibitem[\protect\citeauthoryear{{Hovatta} et~al.,}{{Hovatta}
  et~al.}{2015}]{2015MNRAS.448.3121H}
{Hovatta} T.,  et~al., 2015, \mn@doi [\mnras] {10.1093/mnras/stv220}, \href
  {https://ui.adsabs.harvard.edu/abs/2015MNRAS.448.3121H} {448, 3121}

\bibitem[\protect\citeauthoryear{Ikejiri et~al.,}{Ikejiri
  et~al.}{2011}]{Ikejiri2011}
Ikejiri Y.,  et~al., 2011, \mn@doi [Publications of the Astronomical Society of
  Japan] {10.1093/pasj/63.3.327}, 63, 639

\bibitem[\protect\citeauthoryear{{Kirk}, {Rieger}  \& {Mastichiadis}}{{Kirk}
  et~al.}{1998}]{1998A&A...333..452K}
{Kirk} J.~G.,  {Rieger} F.~M.,   {Mastichiadis} A.,  1998, \aap, \href
  {https://ui.adsabs.harvard.edu/abs/1998A&A...333..452K} {333, 452}

\bibitem[\protect\citeauthoryear{Koide, Shibata  \& Kudoh}{Koide
  et~al.}{1998}]{Koide1998}
Koide S.,  Shibata K.,   Kudoh T.,  1998, \mn@doi [The Astrophysical Journal]
  {10.1086/311204}, 495, L63

\bibitem[\protect\citeauthoryear{{Konigl}}{{Konigl}}{1981}]{Konigl1981}
{Konigl} A.,  1981, \mn@doi [\apj] {10.1086/158638}, \href
  {https://ui.adsabs.harvard.edu/abs/1981ApJ...243..700K} {243, 700}

\bibitem[\protect\citeauthoryear{{MAGIC Collaboration} et~al.,}{{MAGIC
  Collaboration} et~al.}{2020a}]{1es1959}
{MAGIC Collaboration} et~al., 2020a, \mn@doi [\aap]
  {10.1051/0004-6361/201935450}, \href
  {https://ui.adsabs.harvard.edu/abs/2020A&A...638A..14M} {638, A14}

\bibitem[\protect\citeauthoryear{{MAGIC Collaboration} et~al.,}{{MAGIC
  Collaboration} et~al.}{2020b}]{MAGIC2020}
{MAGIC Collaboration} et~al., 2020b, \mn@doi [\aap]
  {10.1051/0004-6361/202037811}, \href
  {https://ui.adsabs.harvard.edu/abs/2020A&A...640A.132M} {640, A132}

\bibitem[\protect\citeauthoryear{{Marscher} \& {Gear}}{{Marscher} \&
  {Gear}}{1985}]{Marsche1985}
{Marscher} A.~P.,  {Gear} W.~K.,  1985, \mn@doi [\apj] {10.1086/163592}, \href
  {https://ui.adsabs.harvard.edu/abs/1985ApJ...298..114M} {298, 114}

\bibitem[\protect\citeauthoryear{{Mo{\'s}cibrodzka} \&
  {Falcke}}{{Mo{\'s}cibrodzka} \& {Falcke}}{2013}]{Monika2013}
{Mo{\'s}cibrodzka} M.,  {Falcke} H.,  2013, \mn@doi [\aap]
  {10.1051/0004-6361/201322692}, \href
  {https://ui.adsabs.harvard.edu/abs/2013A&A...559L...3M} {559, L3}

\bibitem[\protect\citeauthoryear{{Mo{\'s}cibrodzka}, {Falcke}, {Shiokawa}  \&
  {Gammie}}{{Mo{\'s}cibrodzka} et~al.}{2014}]{Monika2014}
{Mo{\'s}cibrodzka} M.,  {Falcke} H.,  {Shiokawa} H.,   {Gammie} C.~F.,  2014,
  \mn@doi [\aap] {10.1051/0004-6361/201424358}, \href
  {https://ui.adsabs.harvard.edu/abs/2014A&A...570A...7M} {570, A7}

\bibitem[\protect\citeauthoryear{{Mo{\'s}cibrodzka}, {Falcke}  \&
  {Shiokawa}}{{Mo{\'s}cibrodzka} et~al.}{2016}]{Monika2016}
{Mo{\'s}cibrodzka} M.,  {Falcke} H.,   {Shiokawa} H.,  2016, \mn@doi [\aap]
  {10.1051/0004-6361/201526630}, \href
  {https://ui.adsabs.harvard.edu/abs/2016A&A...586A..38M} {586, A38}

\bibitem[\protect\citeauthoryear{Mücke, Protheroe, Engel, Rachen  \&
  Stanev}{Mücke et~al.}{2003}]{MUCKE2003}
Mücke A.,  Protheroe R.,  Engel R.,  Rachen J.,   Stanev T.,  2003, \mn@doi
  [Astroparticle Physics] {https://doi.org/10.1016/S0927-6505(02)00185-8}, 18,
  593

\bibitem[\protect\citeauthoryear{{Nilsson} et~al.,}{{Nilsson}
  et~al.}{2018}]{Nilsson2018}
{Nilsson} K.,  et~al., 2018, \mn@doi [\aap] {10.1051/0004-6361/201833621},
  \href {https://ui.adsabs.harvard.edu/abs/2018A&A...620A.185N} {620, A185}

\bibitem[\protect\citeauthoryear{{Padovani} \& {Giommi}}{{Padovani} \&
  {Giommi}}{1995}]{Padovani1995}
{Padovani} P.,  {Giommi} P.,  1995, \mn@doi [\apj] {10.1086/175631}, \href
  {https://ui.adsabs.harvard.edu/abs/1995ApJ...444..567P} {444, 567}

\bibitem[\protect\citeauthoryear{Paliya, Diltz, Böttcher, Stalin  \&
  Buckley}{Paliya et~al.}{2016}]{Paliya_2016}
Paliya V.~S.,  Diltz C.,  Böttcher M.,  Stalin C.~S.,   Buckley D.,  2016,
  \mn@doi [The Astrophysical Journal] {10.3847/0004-637x/817/1/61}, 817, 61

\bibitem[\protect\citeauthoryear{{Patel}, {Shukla}, {Chitnis}, {Dorner},
  {Mannheim}, {Acharya}  \& {Nagare}}{{Patel} et~al.}{2018}]{Patel2018}
{Patel} S.~R.,  {Shukla} A.,  {Chitnis} V.~R.,  {Dorner} D.,  {Mannheim} K.,
  {Acharya} B.~S.,   {Nagare} B.~J.,  2018, \mn@doi [\aap]
  {10.1051/0004-6361/201731987}, \href
  {https://ui.adsabs.harvard.edu/abs/2018A&A...611A..44P} {611, A44}

\bibitem[\protect\citeauthoryear{{Polkas}, {Petropoulou}, {Vasilopoulos},
  {Mastichiadis}, {Urry}, {Coppi}  \& {Bailyn}}{{Polkas}
  et~al.}{2021}]{Polkas2021}
{Polkas} M.,  {Petropoulou} M.,  {Vasilopoulos} G.,  {Mastichiadis} A.,  {Urry}
  C.~M.,  {Coppi} P.,   {Bailyn} C.,  2021, \mn@doi [\mnras]
  {10.1093/mnras/stab1618}, \href
  {https://ui.adsabs.harvard.edu/abs/2021MNRAS.505.6103P} {505, 6103}

\bibitem[\protect\citeauthoryear{{Prince}}{{Prince}}{2019}]{Prince2019a}
{Prince} R.,  2019, \mn@doi [\apj] {10.3847/1538-4357/aaf475}, \href
  {https://ui.adsabs.harvard.edu/abs/2019ApJ...871..101P} {871, 101}

\bibitem[\protect\citeauthoryear{Prince, Raman, Hahn, Gupta  \&
  Majumdar}{Prince et~al.}{2018}]{Prince_2018}
Prince R.,  Raman G.,  Hahn J.,  Gupta N.,   Majumdar P.,  2018, \mn@doi [The
  Astrophysical Journal] {10.3847/1538-4357/aadadb}, 866, 16

\bibitem[\protect\citeauthoryear{Prince, Gupta  \& Nalewajko}{Prince
  et~al.}{2019}]{Prince2019b}
Prince R.,  Gupta N.,   Nalewajko K.,  2019, \mn@doi [The Astrophysical
  Journal] {10.3847/1538-4357/ab3afa}, 883, 137

\bibitem[\protect\citeauthoryear{{Rajput}, {Stalin}, {Sahayanathan}, {Rakshit}
  \& {Mandal}}{{Rajput} et~al.}{2019}]{Rajput2019}
{Rajput} B.,  {Stalin} C.~S.,  {Sahayanathan} S.,  {Rakshit} S.,   {Mandal}
  A.~K.,  2019, \mn@doi [\mnras] {10.1093/mnras/stz941}, \href
  {https://ui.adsabs.harvard.edu/abs/2019MNRAS.486.1781R} {486, 1781}

\bibitem[\protect\citeauthoryear{{Rajput}, {Stalin}  \&
  {Sahayanathan}}{{Rajput} et~al.}{2020}]{Rajput2020}
{Rajput} B.,  {Stalin} C.~S.,   {Sahayanathan} S.,  2020, \mn@doi [\mnras]
  {10.1093/mnras/staa2708}, \href
  {https://ui.adsabs.harvard.edu/abs/2020MNRAS.498.5128R} {498, 5128}

\bibitem[\protect\citeauthoryear{Rani, Krichbaum, Lee, Sokolovsky, Kang, Byun,
  Mosunova  \& Zensus}{Rani et~al.}{2016}]{Rani2017}
Rani B.,  Krichbaum T.~P.,  Lee S.-S.,  Sokolovsky K.,  Kang S.,  Byun D.-Y.,
  Mosunova D.,   Zensus J.~A.,  2016, \mn@doi [Monthly Notices of the Royal
  Astronomical Society] {10.1093/mnras/stw2342}, 464, 418

\bibitem[\protect\citeauthoryear{Rani et~al.,}{Rani et~al.}{2018}]{Rani2018}
Rani B.,  et~al., 2018, \mn@doi [The Astrophysical Journal]
  {10.3847/1538-4357/aab785}, 858, 80

\bibitem[\protect\citeauthoryear{{Rybicki} \& {Lightman}}{{Rybicki} \&
  {Lightman}}{1979}]{Rybicki1979}
{Rybicki} G.~B.,  {Lightman} A.~P.,  1979, {Radiative processes in
  astrophysics}

\bibitem[\protect\citeauthoryear{Safna, Stalin, Rakshit  \& Mathew}{Safna
  et~al.}{2020}]{Safna2020}
Safna P.~Z.,  Stalin C.~S.,  Rakshit S.,   Mathew B.,  2020, \mn@doi [Monthly
  Notices of the Royal Astronomical Society] {10.1093/mnras/staa2622}, 498,
  3578

\bibitem[\protect\citeauthoryear{Schlafly \& Finkbeiner}{Schlafly \&
  Finkbeiner}{2011}]{Schlafly_2011}
Schlafly E.~F.,  Finkbeiner D.~P.,  2011, \mn@doi [The Astrophysical Journal]
  {10.1088/0004-637x/737/2/103}, 737, 103

\bibitem[\protect\citeauthoryear{Schleicher et~al.,}{Schleicher
  et~al.}{2019}]{galaxies7020062}
Schleicher B.,  et~al., 2019, \mn@doi [Galaxies] {10.3390/galaxies7020062}, 7

\bibitem[\protect\citeauthoryear{{Shukla} \& {Mannheim}}{{Shukla} \&
  {Mannheim}}{2020}]{Shukla2020}
{Shukla} A.,  {Mannheim} K.,  2020, \mn@doi [Nature Communications]
  {10.1038/s41467-020-17912-z}, \href
  {https://ui.adsabs.harvard.edu/abs/2020NatCo..11.4176S} {11, 4176}

\bibitem[\protect\citeauthoryear{{Sikora}, {Begelman}  \& {Rees}}{{Sikora}
  et~al.}{1994}]{Sikora1994}
{Sikora} M.,  {Begelman} M.~C.,   {Rees} M.~J.,  1994, \mn@doi [\apj]
  {10.1086/173633}, \href
  {https://ui.adsabs.harvard.edu/abs/1994ApJ...421..153S} {421, 153}

\bibitem[\protect\citeauthoryear{Spitkovsky}{Spitkovsky}{2008}]{Spitkovsky_2008}
Spitkovsky A.,  2008, \mn@doi [The Astrophysical Journal] {10.1086/590248},
  682, L5

\bibitem[\protect\citeauthoryear{{Stickel}, {Padovani}, {Urry}, {Fried}  \&
  {Kuehr}}{{Stickel} et~al.}{1991}]{Stickel1991}
{Stickel} M.,  {Padovani} P.,  {Urry} C.~M.,  {Fried} J.~W.,   {Kuehr} H.,
  1991, \mn@doi [\apj] {10.1086/170133}, \href
  {https://ui.adsabs.harvard.edu/abs/1991ApJ...374..431S} {374, 431}

\bibitem[\protect\citeauthoryear{Thiersen, Zacharias  \& Böttcher}{Thiersen
  et~al.}{2022}]{Thiersen_2022}
Thiersen H.,  Zacharias M.,   Böttcher M.,  2022, \mn@doi [The Astrophysical
  Journal] {10.3847/1538-4357/ac4013}, 925, 177

\bibitem[\protect\citeauthoryear{Ulrich, Maraschi  \& Urry}{Ulrich
  et~al.}{1997}]{Ulrich1997}
Ulrich M.-H.,  Maraschi L.,   Urry C.~M.,  1997, \mn@doi [Annual Review of
  Astronomy and Astrophysics] {10.1146/annurev.astro.35.1.445}, 35, 445

\bibitem[\protect\citeauthoryear{Urry \& Padovani}{Urry \&
  Padovani}{1995}]{Urry_1995}
Urry C.~M.,  Padovani P.,  1995, \mn@doi [Publications of the Astronomical
  Society of the Pacific] {10.1086/133630}, 107, 803

\bibitem[\protect\citeauthoryear{{Urry} et~al.,}{{Urry}
  et~al.}{1997}]{Urry1997}
{Urry} C.~M.,  et~al., 1997, \mn@doi [\apj] {10.1086/304536}, \href
  {https://ui.adsabs.harvard.edu/abs/1997ApJ...486..799U} {486, 799}

\bibitem[\protect\citeauthoryear{{Vaughan}, {Edelson}, {Warwick}  \&
  {Uttley}}{{Vaughan} et~al.}{2003}]{Vaughan_2003}
{Vaughan} S.,  {Edelson} R.,  {Warwick} R.~S.,   {Uttley} P.,  2003, \mn@doi
  [\mnras] {10.1046/j.1365-2966.2003.07042.x}, \href
  {https://ui.adsabs.harvard.edu/abs/2003MNRAS.345.1271V} {345, 1271}

\bibitem[\protect\citeauthoryear{{Weymann}, {Morris}, {Foltz}  \&
  {Hewett}}{{Weymann} et~al.}{1991}]{Weymann1991}
{Weymann} R.~J.,  {Morris} S.~L.,  {Foltz} C.~B.,   {Hewett} P.~C.,  1991,
  \mn@doi [\apj] {10.1086/170020}, \href
  {https://ui.adsabs.harvard.edu/abs/1991ApJ...373...23W} {373, 23}

\bibitem[\protect\citeauthoryear{{Wierzcholska}, {Ostrowski}, {Stawarz},
  {Wagner}  \& {Hauser}}{{Wierzcholska} et~al.}{2015}]{2015A&A...573A..69W}
{Wierzcholska} A.,  {Ostrowski} M.,  {Stawarz} {\L}.,  {Wagner} S.,   {Hauser}
  M.,  2015, \mn@doi [\aap] {10.1051/0004-6361/201423967}, \href
  {https://ui.adsabs.harvard.edu/abs/2015A&A...573A..69W} {573, A69}

\bibitem[\protect\citeauthoryear{{Xue}, {Liu}, {Petropoulou}, {Oikonomou},
  {Wang}, {Wang}  \& {Wang}}{{Xue} et~al.}{2019}]{Xue2019}
{Xue} R.,  {Liu} R.-Y.,  {Petropoulou} M.,  {Oikonomou} F.,  {Wang} Z.-R.,
  {Wang} K.,   {Wang} X.-Y.,  2019, \mn@doi [\apj] {10.3847/1538-4357/ab4b44},
  \href {https://ui.adsabs.harvard.edu/abs/2019ApJ...886...23X} {886, 23}

\bibitem[\protect\citeauthoryear{{de Vaucouleurs}, {de Vaucouleurs}, {Corwin},
  {Buta}, {Paturel}  \& {Fouque}}{{de Vaucouleurs}
  et~al.}{1991}]{Vaucouleurs1991}
{de Vaucouleurs} G.,  {de Vaucouleurs} A.,  {Corwin} Herold~G. J.,  {Buta}
  R.~J.,  {Paturel} G.,   {Fouque} P.,  1991, {Third Reference Catalogue of
  Bright Galaxies}

\makeatother
\end{thebibliography}








\bsp	
\label{lastpage}
\end{document}